# Dynamic CBCT Imaging using Prior Model-Free Spatiotemporal Implicit Neural Representation (PMF-STINR)

Hua-Chieh Shao[1], Tielige Mengke[1], Tinsu Pan[2], You Zhang[1]

*[1]The Medical Artificial Intelligence and Automation (MAIA) Laboratory*
*Department of Radiation Oncology, University of Texas Southwestern Medical Center, Dallas, TX 75390, USA*
*[2]Department of Imaging Physics*
*University of Texas MD Anderson Cancer Center, Houston, TX, 77030, USA*

## Abstract

**Objective:** Dynamic cone-beam computed tomography (CBCT) can capture high-spatial-resolution, time-varying images for motion monitoring, patient setup, and adaptive planning of radiotherapy. However, dynamic CBCT reconstruction is an extremely ill-posed spatiotemporal inverse problem, as each CBCT volume in the dynamic sequence is only captured by one or a few X-ray projections, due to the slow gantry rotation speed and the fast anatomical motion (e.g., breathing).

**Approach:** We developed a machine learning-based technique, prior-model-free spatiotemporal implicit neural representation (PMF-STINR), to reconstruct dynamic CBCTs from sequentially acquired X-ray projections. PMF-STINR employs a joint image reconstruction and registration approach to address the under-sampling challenge, enabling dynamic CBCT reconstruction from singular X-ray projections. Specifically, PMF-STINR uses spatial implicit neural representation to reconstruct a reference CBCT volume, and it applies temporal INR to represent the intra-scan dynamic motion with respect to the reference CBCT to yield dynamic CBCTs. PMF-STINR couples the temporal INR with a learning-based B-spline motion model to capture time-varying deformable motion during the reconstruction. Compared with the previous methods, the spatial INR, the temporal INR, and the B-spline model of PMF-STINR are all learned on the fly during reconstruction in a one-shot fashion, without using any patient-specific prior knowledge or motion sorting/binning.

**Main results:** PMF-STINR was evaluated via digital phantom simulations, physical phantom measurements, and a multi-institutional patient dataset featuring various imaging protocols (half-fan/full-fan, full sampling/sparse sampling, different energy and mAs settings, etc.). The results showed that the one-shot learning-based PMF-STINR can accurately and robustly reconstruct dynamic CBCTs and capture highly irregular motion with high temporal (~0.1s) resolution and sub-millimeter accuracy.

**Significance:** PMF-STINR can reconstruct dynamic CBCTs and solve the intra-scan motion from conventional 3D CBCT scans without using any prior anatomical/motion model or motion sorting/binning. It can be a promising tool for motion management by offering richer motion information than traditional 4D-CBCTs.

## 1. Introduction

*1.1 Background and study overview*

Cone-beam computed tomography (CBCT) is widely used in clinical practice. In radiotherapy, CBCT provides high-spatial-resolution volumetric imaging guidance for treatment setup, dose verification, and adaptive therapy (Jaffray *et al.*, 2002; Oldham *et al.*, 2005). For CBCT imaging, cone-beam projections are acquired by a source-detector pair that rotates around the patient. The acquisition efficiency is limited by the rotation speed which is generally restricted to ~6°/second (s) for patient safety. Accordingly, it takes ~1 minute (min) or more to acquire a 360° scan. Due to the long acquisition time, patient anatomical motion, mostly respiration (3-5 s per breathing cycle), results in artifacts and blurriness in the reconstructed CBCTs (Rit *et al.*, 2011). To address the artifacts and resolve the underlying motion, four-dimensional (4D)-CBCT was developed as the current clinical state-of-the-art (Sonke *et al.*, 2005; Zhang *et al.*, 2013b; Abulimiti *et al.*, 2023). 4D-CBCT sorts the projections into a pre-defined set of motion bins and stacks semi-static CBCTs reconstructed from each bin to represent an averaged motion pattern. The motion sorting assumes that the underlying anatomical motion is periodic and regular, which is in general false (Yasue *et al.*, 2022). Correspondingly, 4D-CBCT cannot capture time-resolved irregular motion which may significantly impact patient setup and dose delivery accuracy (Li *et al.*, 2018; Vergalasova *et al.*, 2011; Clements *et al.*, 2013). Moreover, 4D motion sorting is usually based on surrogates (e.g., surface reflective markers), and can be inaccurate due to limited surrogate-anatomy motion correlation (Yan *et al.*, 2008).

A fundamental solution to the limitations of 4D-CBCT is to reconstruct a dynamic sequence of CBCTs (one CBCT for one projection), which eliminates the uncertainties from motion sorting to capture both regular and irregular motion. Dynamic CBCTs can reveal patient anatomy variations with the utmost spatial and temporal resolutions. For radiotherapy, in the pre-treatment stage, dynamic CBCTs can be used to capture the true range of motion to optimize the treatment margin and select the most appropriate motion management technique. During the treatment, the dynamic CBCT can be coupled with the dose delivery sequence to reconstruct dynamic doses and reveal true 'accumulated' dosage (Zou *et al.*, 2014). Such information can then be applied to guide the adaptation of future treatments to preserve the dose delivery precision (Brock, 2019). However, dynamic CBCT is not yet clinically available, mostly due to the lack of a robust technique to reconstruct CBCTs from singular X-ray projections. The information captured by a single 2D projection is extremely limited for CBCT reconstruction, as conventional reconstruction methods require hundreds of projections (Feldkamp *et al.*, 1984). Although there are numerous studies on 4D-CBCT imaging and reconstruction (Shieh *et al.*, 2019; Zhi *et al.*, 2020; Huang *et al.*, 2020; Zhang *et al.*, 2013b; Zhang *et al.*, 2016; Wang and Gu, 2013; Bergner *et al.*, 2009; Yan *et al.*, 2014), the corresponding studies on dynamic CBCT reconstruction are very scarce (Cai *et al.*, 2014; Gao *et al.*, 2018; Jailin *et al.*, 2021; Zhang *et al.*, 2023). The few available methods are either limited to preliminary studies based on simplified non-cone-beam geometries; or relying on specifically crafted, prior anatomical/motion models that are susceptible to differing CBCT acquisition conditions and patient anatomy/motion variations. Most of the studies are evaluated on simulated data rather than real clinical cone-beam projections, which did not fully demonstrate their clinical applicability and translation potential.

Our proposed study marks the first general, prior model-free dynamic CBCT reconstruction approach to address the above-mentioned challenges. Specifically, we have made the following contributions:

- We developed a one-shot dynamic CBCT reconstruction technique that can reconstruct hundreds of motion-resolved CBCT volumes from a conventional 3D CBCT scan, without using any prior modeling or motion binning/sorting. The developed technique offers a clear advantage over the current 4D-CBCT approach and can reveal both regular and irregular volumetric motion to better inform treatment strategies and decisions.
- To the best of our knowledge, our study marks the first comprehensive evaluation and validation of a dynamic CBCT reconstruction technique. We evaluated our approach using a digital phantom simulation study, a dynamic thorax phantom measurement study, and a patient study using multi-institutional clinical cone-beam projections featuring combined anatomical, motion, and imaging variations.

## *1.2 Related works*

Dynamic CBCT reconstruction has been a challenging problem with few reported studies. Cai et al. (Cai *et al.*, 2014) proposed a low-rank reconstruction approach, assuming the dynamic CBCTs can be approximated as a linearly-weighted sequence of 20 basis images derived from the projection data. The proof-of-concept study showed encouraging results for time-resolved image reconstruction, but the algorithm was only evaluated for the 2D fan-beam geometry, using simulated data from a digital phantom under regular motion. Gao et al. (Gao *et al.*, 2018) proposed to use a motion model learned from a fixed-angle fluoroscopy sequence to inform the reconstruction of semi-dynamic CBCTs via a low-rank approach. The method does not reconstruct a CBCT per X-ray projection and requires additional projection sorting/binning, and is susceptible to the uncertainties of the prior motion model learned from the fluoroscopy sequence. Its evaluation set is limited to digital and physical phantoms as well. In comparison to these reconstruction-based approaches, a series of studies based on deformation-driven approaches also attempted 4D or dynamic CBCT reconstruction by infusing assumptions including prior knowledge of the patient anatomy (anatomical model) and/or the motion (motion model). With prior knowledge of the patient anatomy (e.g., a prior CT), techniques were developed to use under-sampled projections to deform the prior CT to 4D-CBCTs, assuming the underlying anatomical model does not change (Zhang *et al.*, 2013a; Dang *et al.*, 2015). To enable single projection-based dynamic CBCT reconstruction, a principal component analysis (PCA)-based motion model derived from patient-specific prior 4D-CTs was introduced, to achieve further dimension reduction to satisfy the extreme under-sampling scenario (Li *et al.*, 2011; Wei *et al.*, 2020). However, the assumption of an invariant anatomical model may be invalidated by non-deformation-related anatomical changes over time, for instance, contrast enhancement, normal tissue inflammation, and disease progression (Zhang *et al.*, 2017). The use of an anatomical model derived from a different machine rather than the same CBCT device also poses additional challenges caused by differing energy/scatter/noise conditions and image intensity variations (Zhang *et al.*, 2015). The assumption of an invariant motion model, on the other hand, may not account for inter-fractional deformation and motion pattern variations (Zhang *et al.*, 2013a). The venue to generate such a motion model (e.g., prior 4D-CTs), may contain motion sorting artifacts or not be available for some patients. To use the prior motion model, a co-registration is usually needed to align the prior and new imaging coordinate systems, which can introduce additional errors. In addition to the above-mentioned pure reconstruction-driven or deformation-driven approaches, Jailin et al. (Jailin *et al.*, 2021) used joint image reconstruction and registration for dynamic CBCT imaging. They combined image reconstruction with a mesh-based registration model, where the temporal motion kinematics were initialized as pre-defined functions or tracked surrogate curves. Although a certain degree of relaxation is allowed to correct errors within the initial temporal kinematics, it is uncertain to which extent the correction can be made in case of large motion mismatches. The method was evaluated by a single CBCT scan, and the evaluation was focused on the motion blurriness reduction of the reconstructed CBCT rather than accuracy of the tracked motion trajectory.

In recent years, deep learning (DL)-based approaches were also investigated for severely under-sampled CT/CBCT reconstruction. Shen et al. (Shen *et al.*, 2019) developed a patient-specific encoder-decoder network to directly map an X-ray projection to a 3D volume. However, the projection-to-CBCT inverse mapping can be extremely ill-conditioned, as such models heavily rely on the comprehensiveness of the training dataset. A small out-of-distribution variation, for instance the anatomy/motion change or the imaging parameter change, may result in large instabilities from these models. In addition to direct reconstruction, deformable registration-based DL methods were also developed (Wei *et al.*, 2020; Shao *et al.*, 2022; Wei *et al.*, 2019). Similar to traditional pure deformable registration-based methods, these techniques are susceptible to the uncertainties from non-deformation-induced anatomy changes and imaging system mismatches. These models also need to be trained with a large dataset with simulated motion scenarios, where the comprehensiveness of the training dataset determines their accuracy. In addition, the reconstruction- or registration-based DL models are usually trained and may only work under a specific gantry angle, although recent studies have successfully achieved angle-agnostic inference with additional geometry priors and ancillary information like optical imaging (Shao *et al.*, 2023a; Shao *et al.*, 2023b).

Recently, a machine learning technique named implicit neural representation (INR) has emerged for imaging applications (Mildenhall *et al.*, 2021). INR uses the non-parametric representation capability of neural networks to learn implicit mapping of complex 3D scenes (e.g., CBCTs) from sparse 2D views (e.g., cone-beam projections) (Mildenhall *et al.*, 2022). Acting as a universal function approximator, INR takes geometric coordinates of a scene as inputs and maps them to queried physical features at the coordinates (e.g., linear attenuation coefficients of CBCTs). Compared with conventional voxel-grid-based representation, INR can take non-integer coordinates as inputs for continuous mapping, allowing resolution-agnostic representation of the underlying scene. The continuous, differentiable, and implicit nature of INR allows it to map highly complex medical images in a compact format and promotes image reconstruction from under-sampled signals. Shen et al. (Shen *et al.*, 2022) developed a novel framework that applies the INR to encode a prior anatomical image, and then uses sparsely-sampled on-board projections to update the INR to achieve limited-sampling-based reconstruction. Such a technique, using a prior anatomical model, can be affected by the imaging system/protocol mismatches between the prior and new acquisitions. Another study by Wu et al. (Wu *et al.*, 2023) developed a self-supervised coordinate projection network (SCOPE) for sparse-view CT reconstruction. SCOPE uses INR as an image continuity prior to constrain the solution space, and combines INR-projected dense-view sinogram and the original sparse-view sinogram for final CT reconstruction. However, SCOPE is limited to sparse-view CT reconstruction with 60-120 parallel or fan-beam rays, and its application to single projection-driven dynamic CBCT reconstruction is expected to be challenging. Other works, including GRFF (Tancik *et al.*, 2020a), IntraTomo (Zang *et al.*, 2021), and CoIL (Sun *et al.*, 2021), are similarly developed for sparse-view CT reconstruction rather than dynamic CBCT imaging, with the latter being a significantly more challenging problem. Another recent study, by Reed et al. (Reed *et al.*, 2021), used INR and polynomial-based temporal motion fields fitting to reconstruct limited-angle dynamic CT images for the parallel-beam geometry (INR-poly). INR-poly assumes the intra-scan motion of CT imaging can be approximated by polynomial-fitted motion fields. However, the polynomial fitting limits the potential of INR-poly to describe complex motion, and the motion has to be slow related to the scan speed so that the subject/anatomy remains static within each limited-angle CT scan. The imaging arm of CBCT for radiotherapy rotates much slower than a diagnostic CT, and such assumption will not stand. Inspired by the potential of INR, we recently developed an INR-based joint deformation and reconstruction framework, spatial and temporal implicit neural representation (STINR), for dynamic CBCT reconstruction (Zhang *et al.*, 2023). Compared with previous DL-based approaches that require a large training dataset, STINR reconstructs dynamic CBCTs through directly learning from scan-specific cone-beam projections ('one-shot'), avoiding potential issues of overfitting and domain shift. By STINR, we decoupled the spatiotemporal dynamic CBCT inverse problem into reconstructing a reference CBCT volume (spatial INR) and the intra-scan motion (temporal INR) related to the reference CBCT, with the help of a prior patient-specific motion model derived from a separate 4D-CT scan. STINR has shown substantially improved dynamic CBCT and motion reconstruction accuracy than other methods, as the combination of patient-specific motion modeling and temporal INR allows the description of highly-irregular and complex motion patterns. However, STINR suffers from two major disadvantages: (1). It requires a pre-acquired high-quality 4D imaging set to extract a motion model, which may not be always available; and the motion model learned from it can be outdated and error-prone for following dynamic CBCT reconstruction; and (2). Prior to the one-shot reconstruction by STINR, a 4D sorting/binning procedure is still required to reconstruct a phase-specific, initial CBCT volume to fit the prior motion model. In addition, the evaluation of the STINR study is limited to a digital phantom study and a patient simulation study. The patient simulation study uses a patient 4D-CBCT set and its derivative PCA motion model to simulate dynamic CBCTs and corresponding cone-beam projections for dynamic reconstruction. Thus the same motion model is shared between the dynamic CBCT simulation and the STINR reconstruction, which is overly ideal in real clinical scenarios. The projections were simulated large enough to cover the full anatomy in the field-of-view, beyond what can be acquired by current clinical systems. Actual clinical acquisitions, using realistic projection sizes and a diverse set of imaging protocols, are warranted to evaluate the dynamic CBCT imaging framework in capturing clinically-observed motion.

Built on the foundation of STINR (Zhang *et al.*, 2023), in this study we developed a prior model-free STINR (PMF-STINR) framework to solve the above-mentioned challenges in dynamic CBCT imaging and address the disadvantages of STINR. Compared with the original STINR (and the other methods), PMF-STINR does not use any prior anatomical or motion model, thus is not prone to the limitations of strong *a priori* assumptions. Instead, it uses a data-driven motion model directly learned on the fly from the acquired cone-beam projections. Moreover, PMF-STINR does not require any motion sorting/binning of the cone-beam projections, in contrast to the original STINR framework which still needed motion binning in the early stages of reconstruction. Compared with the original STINR framework, PMF-STINR also introduced multi-resolution hash encoding and spatiotemporal regularizations to further improve the reconstruction speed and accuracy, and refined the training strategy to address the spatiotemporal ambiguity and under-sampling challenges. PMF-STINR was evaluated via combined phantom simulation, phantom measurement, and patient measurement studies to comprehensively assess its clinical translation potential. For the patient study, a motion evaluation strategy using semi-automatically or automatically-tracked anatomical landmarks on the 2D projection domain was employed under the lack of 'ground-truth' 3D reference.

## 2. Materials and Methods

### 2.1 Dynamic CBCT reconstruction problem overview

For CBCT imaging, define $\{\boldsymbol{p}_t\}_{t=0}^{N_p-1}$ as a consecutive sequence of cone-beam X-ray projections of a dynamic subject, where $t$ denotes the frame index labeling the acquisition order of each projection, and $N_p$ is the total number of projections. Dynamic CBCT imaging is to reconstruct a sequence of 3D volumes $\{\boldsymbol{I}(\boldsymbol{x},t)\}_{t=0}^{N_p-1}$ from $\{\boldsymbol{p}_t\}_{t=0}^{N_p-1}$ to represent the dynamic subject. Physically, $\{\boldsymbol{I}(\boldsymbol{x},t)\}_{t=0}^{N_p-1}$ represents the linear attenuation coefficients at spatial coordinates $\boldsymbol{x} \in \mathbb{R}^3$ and the temporal frame index $t \in \mathbb{R}$. The reconstruction problem is typically solved within an optimization framework:

$$\{\hat{\boldsymbol{I}}(\boldsymbol{x},t)\} = \underset{\{\boldsymbol{I}(\boldsymbol{x},t)\}}{\operatorname{argmin}}(\|\mathcal{P}\{\boldsymbol{I}(\boldsymbol{x},t)\} - \{\boldsymbol{p}_t\}\|^2 + \lambda\,R)\,, \tag{1}$$

where $\mathcal{P}$ denotes the projection matrix, and $\lambda$ is the weighting factor for an regularization term $R$. Note that the superscripts and subscripts are removed outside of the brackets of $\{I(x, t)\}$ and $\{p_t\}$ in the equation for simplicity, which do not change the denotations. The first term of Eq. 1 enforces the projection-domain data fidelity, while $R$ regularizes the image/motion in a transformed domain to avoid overfitting and sub-optimal solutions, e.g., total variation (Zhang *et al.*, 2017).

Solving the optimization problem of Eq. 1 can be extremely challenging. A whole dynamic sequence $\{I(x, t)\}_{t=0}^{N_p - 1}$ contains $\mathcal{O}(10^8)$ or more voxels to solve, while the volumetric information at each moment $t$ is only captured by a single 2D projection $p_t$. However, assuming the underlying anatomy remains unchanged during the scan (which is generally valid under physiological motion), we could use a joint reconstruction and deformable registration approach to simplify the inverse problem. The joint approach de-couples the spatiotemporal inverse problem into reconstructing a reference-frame CBCT $I_{ref}(x)$ and solving intra-scan motion with respect to $I_{ref}(x)$, which can be described as a sequence of time-dependent deformation vector fields (DVF) $\{d(x, t)\}_{t=0}^{N_p - 1}$. The whole dynamic CBCT sequence $\{I(x, t)\}_{t=0}^{N_p - 1}$ can be obtained by deforming/warping the reference CBCT with $\{d(x, t)\}_{t=0}^{N_p - 1}$:

$$\{I(x, t)\} = I_{ref}(x + \{d(x, t)\}) . \tag{2}$$

To reduce the dimensionality of the solution space for $\{d(x, t)\}_{t=0}^{N_p - 1}$, the inherent redundancy of anatomic motion could be further leveraged to yield a low-rank representation of $\{d(x, t)\}_{t=0}^{N_p - 1}$. Each time-dependent motion field $d(x, t)$ is approximately separable (Zhao *et al.*, 2012) as a summation of products of spatial ($e_i(x)$) and temporal ($w_i(t)$) components:

$$d(x, t) = \sum_{i=1}^{L} w_i(t) \times e_i(x) \tag{3}$$

These spatial components $\{e_i(x)\}_{i=1}^{L}$ maximally capture the motion variations, and compose a basis set (motion basis components, MBC) to explain the motion space. Previous studies, including STINR, use prior 4D images like 4D-CTs to extract principal motion components by principal component analysis, and those principal motion components are conceptually-equivalent to the MBCs here. However, the PMF-STINR approach of this study directly learned MBCs from the acquired projection data, without using any prior information. Considering Eqs. 1-3, dynamic CBCT imaging is equivalent to reconstructing a reference CBCT $I_{ref}$, while finding time-varying linear weightings ($\{w_i(t)\}_{t=0, i=1}^{N_p-1, L}$) of MBCs ($\{e_i(x)\}_{i=1}^{L}$) that maximally account for the underlying anatomical motion. The dimension reduction of unknowns achieved by the joint reconstruction and deformation (Eqs. 1 and 2), and the further spatiotemporal decoupling of the DVFs (Eq. 3), render the dynamic CBCT reconstruction a significantly more tractable problem. In addition, for PMF-STINR, we also employed a B-spline-based parametrization of $\{e_i(x)\}_{i=1}^{L}$ to further reduce the number of unknowns to $\mathcal{O}(10^4)$ (Sec. 2.2.3). As previous studies have shown that three principal motion components are sufficient to represent the complex breathing motion, for PMF-STINR we also used 3 MBCs for each Cartesian direction ($L = 3$), yielding $\{e_i(x)\}_{i=1}^{3}$ (Li *et al.*, 2011). Below we introduced the details of PMF-STINR, including the general framework and each of its components.

*2.2 The PMF-STINR method*

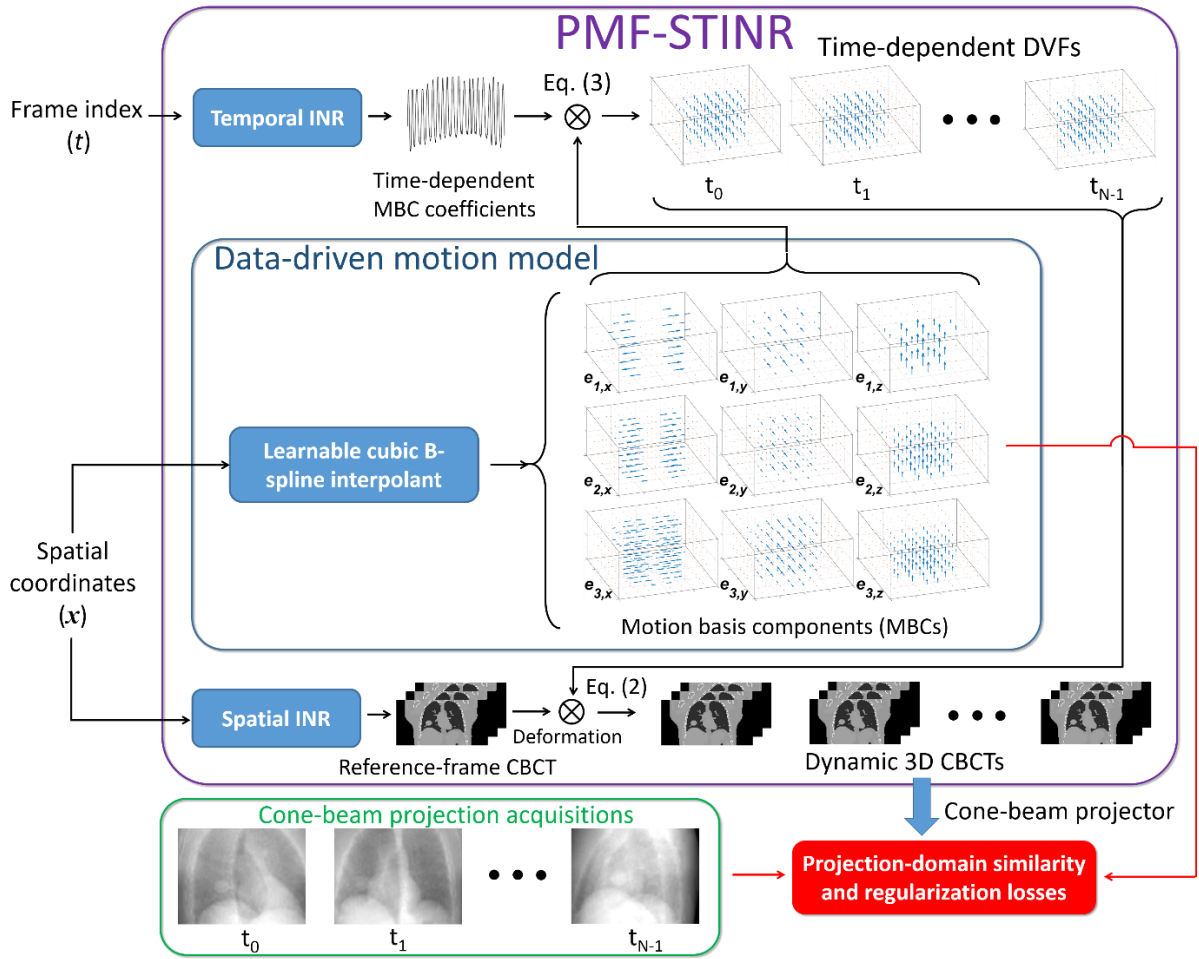

**Fig. 1** Workflow of PMF-STINR. PMF-STINR solves a sequence of dynamic CBCTs, by using a spatial INR to reconstruct a reference-frame CBCT and a temporal INR as well as a data-driven B-spline-based motion model to represent the intra-scan motion.

Figure 1 illustrates the workflow of PMF-STINR, which comprises three major blocks: spatial INR, temporal INR, and a learnable cubic B-spline interpolant for data-driven motion modeling. The spatial INR reconstructs a reference-frame CBCT of the dynamic sequence, and the temporal INR, in conjunction with the data-driven B-spline motion model, solves the time-dependent DVFs of the subject with respect to the reference CBCT. The motion model learns data-driven MBCs $\{e_i(x)\}_{i=1}^3$ from the cone-beam projections, while the temporal INR captures the time-varying temporal coefficients $\{w_i(t)\}_{t=0,i=1}^{N_p-1,3}$ for the MBCs. PMF-STINR uses B-splines to parametrize the dense MBCs with coarser grids of control points than those of voxels, to leverage the piecewise smooth nature of motion fields. The values of control points, representing the motion patterns, are learned via the PMF-STINR framework.

For PMF-STINR, the spatial and temporal INRs and the data-driven motion model are jointly trained, reconstructing the reference-frame CBCT and solving the intra-scan motion simultaneously. The network training is driven by both projection-domain data fidelity loss and spatiotemporal regularization losses. In the following subsections we described each component of PMF-STINR and the training strategy.

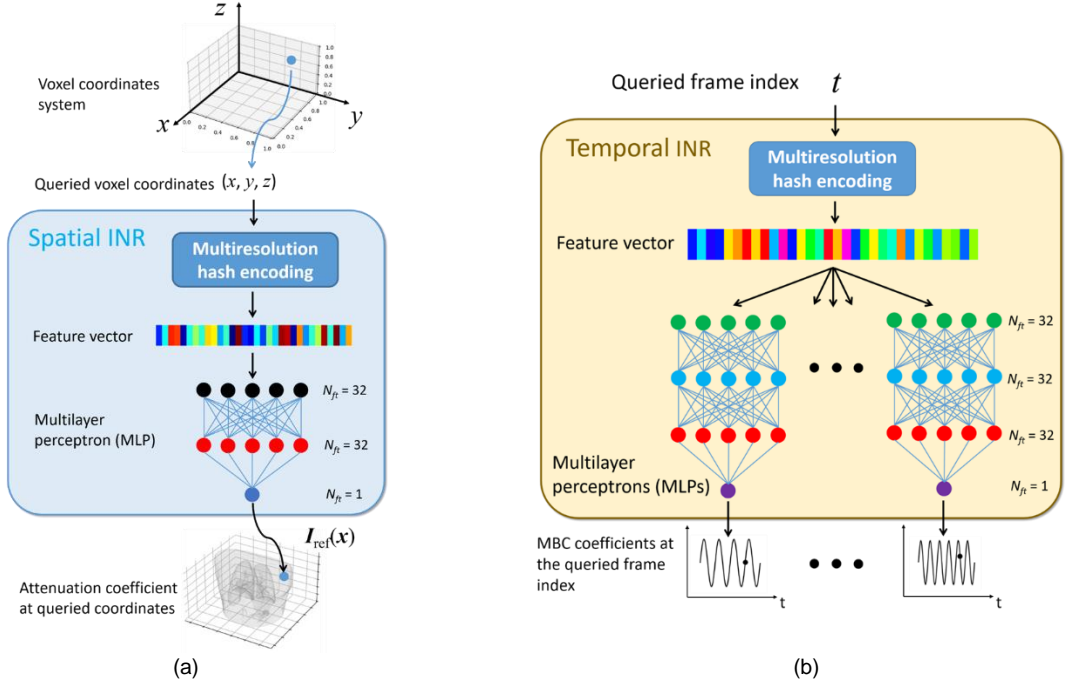

**Fig. 2** Details of the spatial and temporal INRs. The spatial INR reconstructs a reference-frame CBCT by learning a continuous mapping from the voxel coordinate system to the linear attenuation coefficient, utilizing multiresolution hash encoding and a multilayer perceptron (MLP). The temporal INR maps a frame index $t$ to the temporal coefficient $\boldsymbol{w}_i(t)$ of the MBCs, using multiresolution hash encoding and nine MLPs. Each of the nine MLPs corresponds to a Cartesian component of $\{\boldsymbol{e}_i(\boldsymbol{x})\}_{i=1}^3$.

Figure 2 presents the details of the spatial INR (Fig. 2a) and the temporal INR (Fig. 2b). The spatial INR takes a voxel coordinate $\boldsymbol{x}$ of the reference CBCT as input, and maps $\boldsymbol{x}$ to the attenuation coefficient $\boldsymbol{I}_{ref}(\boldsymbol{x})$. Correspondingly, the temporal INR captures the time-varying temporal coefficients to represent intra-scan motion. Its input is a frame index $t$, and the outputs are the time-dependent coefficients $\boldsymbol{w}_i(t)$ ($i$ = 1-3) of the MBCs. Before inputting into the INRs, the voxel coordinate $\boldsymbol{x}$ or the frame index $t$ is rescaled to [-1, 1]. By the spatial INR, the whole reference CBCT can be queried by traversing all voxel coordinates. Similarly, the whole temporal sequence can be obtained from the temporal INR by traversing all frame indices of the acquisition.

The spatial INR comprises a learnable multiresolution hash encoding step and a multilayer perceptron (MLP) that acts as a non-parametric universal function approximator. In contrast, the temporal INR comprises a hash encoding step and nine MLPs. Each of the nine MLPs corresponds to a Cartesian component ($x$, $y$, $z$) of the three MBCs $\{\boldsymbol{e}_i(\boldsymbol{x})\}_{i=1}^3$, and all MLPs use the same network architecture. As MLP alone is ineffective in learning high-frequency details (Tancik *et al.*, 2020b), the added hash encoding step of the spatial or the temporal INR helps to facilitate the fine detail learning (Muller *et al.*, 2022). The hash encoding maps a queried input ($\boldsymbol{x}$ or $t$) to a high-dimensional feature vector, via a multilevel encoding scheme. At each level, a uniform grid of points is set up (denser grids for higher levels), and a predefined hash function (Muller *et al.*, 2022) maps the neighboring grid points of the queried input ($\boldsymbol{x}$ or $t$) to the indices of a hash table to retrieve their corresponding feature vectors. Based on the feature vectors of neighboring grids, the feature vector of the queried input is derived by bi- and tri-linear interpolations for the temporal and spatial INRs, respectively. Finally, the feature vectors of the queried input at all resolution levels are concatenated and fed into the following MLP. The hash table entries are learnable parameters to handle hash collisions and extract most important feature values (Muller *et al.*, 2022). Compared with the Fourier frequency encoding used in the original STINR framework (Zhang *et al.*, 2023), the hash encoding improves the learning efficiency and reduces the needed complexity of the subsequent MLP to map subjects/motion. The hyper-parameters of the hash encoding were adopted from those recommended by Muller et. al (Muller *et al.*, 2022).

With hash encoding, the MLP of the spatial INR only used three fully connected layers. The feature numbers of the input, hidden, and output layers were 32, 32, and 1, respectively. The periodic activation function, SIREN (Sitzmann *et al.*, 2020), was used to capture fine details of CBCT. Correspondingly, the MLPs of the temporal INR comprised one input and one output layer, and two hidden layers. The feature numbers were 32 for the input and hidden layers, and 1 for the output layer.

### 2.2.3 Data-driven motion modeling

Coupled with the temporal INR (Sec. 2.2.2), the data-driven motion model of PMF-STINR learns the MBCs $\{e_i(x)\}_{i=1}^3$ of the dynamic DVFs directly from the cone-beam projections, without using any patient-specific prior knowledge. As mentioned in Sec. 2.1, $\{e_i(x)\}_{i=1}^3$ of PMF-STINR are represented by cubic B-spline interpolants, which parametrize the 3D Cartesian space via B-spline interpolations between a uniform 3D grid of control points. The interpolant takes a form of cubic polynomials, and the first derivative of the interpolant is smooth across the joints of the cubic splines. The global continuity and smoothness are considered desired properties of DVFs, preventing self-folding of soft tissues and preserving topology. The grid parametrization is computationally efficient and allows dimension reduction from the original voxel representations, while maintaining flexibility with local control of the DVFs. In addition, the B-spline-based interpolants are numerically stable.

The anatomical motion, especially respiration, usually involves deformations across multiple scales. For example, the tissue deformation caused by the diaphragm contraction is typically bulky and large-scale, while the lung nodule motion is more local. To better account for the complexity of motion, PMF-STINR applies a hierarchical multiresolution strategy to represent the motion fields, with similar approaches demonstrated effective in deformable registration to avoid sub-optimal solutions (Lester and Arridge, 1999). By the multiresolution strategy (Fig. 1), the three MBCs for each Cartesian direction are deliberately assigned to three different spatial resolutions and motion scales, with each being represented by a B-spline interpolant of a different grid resolution.

Specifically, let $\left\{(s_{i,l}^x, s_{i,l}^y, s_{i,l}^z)\right\}_{l=0}^{N_i}$ denote the coordinates of 3D grid points for MBCs $e_i(x)$ of a specific resolution level, where $N_i$ is the number of grid points along a Cartesian direction (same for all three Cartesian directions). Let $e_{i,k}$ be the $k^{\text{th}}$ Cartesian directional component of $e_i(x)$. The 3D B-spline interpolation of $e_{i,k}(x, y, z)$ at a queried point $(x, y, z)$ is performed via sequential 1D interpolations across the $x$, $y$, and $z$ directions. For instance, when the B-spline interpolation was performed in the $x$-direction, $e_{i,k}$ was written as a linear superposition of cubic B-spline basis functions:

$$e_{i,k}\left(x, s_{i,l'}^y, s_{i,l''}^z\right) = \sum_l P_{i,k}(l, l', l'') \times B_{i,l,3}(x), \qquad (4)$$

where $l$ denotes all grid points along the $x$ direction, and $l'$ and $l''$ are the neighboring grid points of queried $y$ and $z$. $P_{i,k}(l, l', l'')$ denotes the value of the control point at grid point $(s_{i,l}^x, s_{i,l'}^y, s_{i,l''}^z)$, and $B_{i,l,3}(x)$ is the cubic basis function which can be derived by the Cox-de Boor recursion formula:

$$B_{i,l,0}(x) = \begin{cases} 1 & \text{if } s_{i,l}^x \leq x < s_{i,l+1}^x \\ 0 & \text{otherwise} \end{cases}, \tag{5}$$

$$B_{i,l,p}(x) = \frac{x - s_{i,l}^x}{s_{i,l+p}^x - s_{i,l}^x} B_{i,l,p-1}(x) + \frac{s_{i,l+p+1}^x - x}{s_{i,l+p+1}^x - s_{i,l+1}^x} B_{i,l+1,p-1}(x). \tag{6}$$

Here $p$ denotes the B-spline order. After the $x$-direction interpolation, the $y$- and $z$-direction interpolations were performed sequentially in a similar manner. By PMF-STINR, the values of the control points $\{P_{l,k}\}$ in Eq. 4 are learnable parameters that characterize the underlying motion. In this study, we adopted the B-spline interpolant model from (Tegunov and Cramer, 2019). The input into the B-spline interpolant model was the voxel coordinates (normalized to [0, 1]), and the model output the vector fields $\boldsymbol{e}_i(\boldsymbol{x})$ at the queried coordinates.

With the MBCs $\boldsymbol{e}_i(\boldsymbol{x})$ of each specific resolution levels learned by PMF-STINR, the MBCs of all three resolution levels were weighted and summed by the related coefficients $\boldsymbol{w}_i(t)$ that are simultaneously learned with the temporal INR (Sec. 2.2.2), to capture both global and local motion. The resulting temporal motion fields, dynamic DVFs, can deform the reference CBCT volume $\boldsymbol{I}_{ref}(\boldsymbol{x})$ obtained from the spatial INR (Sec. 2.2.2) to yield dynamic CBCT volumes. While different strategies have been introduced to mitigate the ill-posed spatiotemporal reconstruction problem, training the spatial and temporal INRs and the data-driven, B-spline-based motion model simultaneously can be challenging, due to the spatiotemporal ambiguity and the projection under-sampling. To address this issue, a three-staged, progressive training strategy was developed for PMF-STINR and introduced below.

### 2.2.4 The progressive training strategy of PMF-STINR

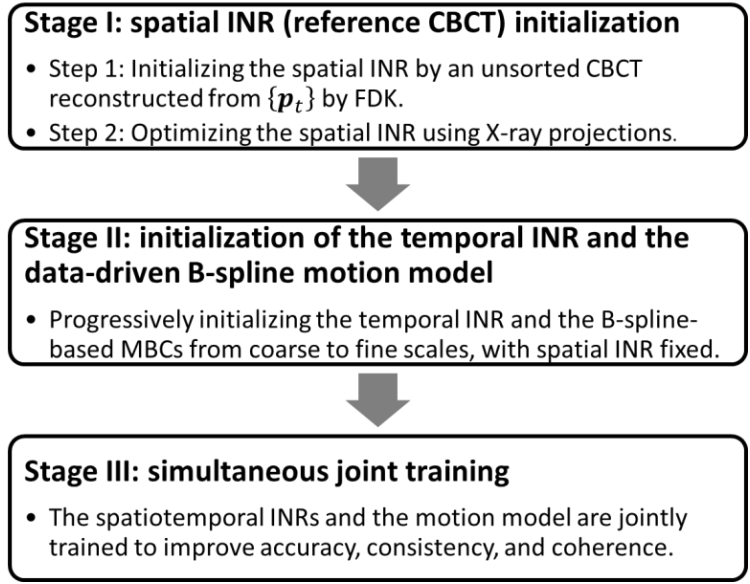

**Fig. 3** The three-staged progressive training strategy of PMF-STINR.

Built on the foundation of the prior STINR framework (Zhang *et al.*, 2023), PMF-STINR uses a progressive three-staged training strategy to initialize different components before simultaneous training (Fig. 3). The strategy progressively increases the learning complexity through three stages to help avoid local optima during the training. In Stage I, the spatial INR was initialized by a motion-averaged CBCT reconstructed from all available projections $\{\boldsymbol{p}_t\}_{t=0}^{N_p-1}$. The approximate reference CBCT $\boldsymbol{I}_{approx}(\boldsymbol{x})$ was reconstructed using the Feldkamp-Davis-Kress (FDK) algorithm (Feldkamp *et al.*, 1984). The fidelity loss of this step ($L_{img}$) was thus defined in the image domain:

$$L_{img} = \frac{1}{N_{voxel}} \sum_{i=1}^{N_{voxel}} \left\| INR_{spa}(\boldsymbol{x_i}) - \boldsymbol{I}_{approx}(\boldsymbol{x_i}) \right\|^2, \tag{7}$$

where $N_{voxel}$ is the total number of voxels in the reference-frame CBCT, and $INR_{spa}$ denotes the spatial INR. As the full-projection reconstruction contains FDK-related artifacts, in the second step of Stage I, $INR_{spa}$ was instead optimized through a fidelity loss ($L_{prj}^a$) defined in the projection domain, similar to the conventional iterative forward-backward projection:

$$L_{prj}^a = \frac{1}{N_{pixel}} \sum_{j=1}^{N_{pixel}} \left\| \mathcal{P} \, INR_{spa}(\boldsymbol{x}) - \{\boldsymbol{p}_t\} \right\|^2, \tag{8}$$

where $N_{pixel}$ is the number of projection pixels. In addition to the fidelity loss, an image-domain L-1 regularization loss (total variation, TV) was also introduced in this step to promote the sparsity of the reference CBCT in its gradient domain:

$$L_{TV} = \frac{1}{N_{voxel}} \sum_i \left| \nabla \boldsymbol{I}_{ref}(\boldsymbol{x_i}) \right|, \tag{9}$$

where $\nabla$ denotes the gradient operator. The overall loss function for the second step of Stage I is then:

$$L_{tot}^I = L_{prj}^a + \lambda_{TV} \, L_{TV}. \tag{10}$$

The value of $\lambda_{TV}$ was set to $1 \times 10^{-3}$ via empirical searching, and the same value was used throughout the following Stages (II and III). The training epochs of the first and second steps of Stage I were 1,000 and 600, respectively.

In Stage II, the temporal INR and the data-driven B-spline motion model were initialized, with respect to the reference-frame CBCT $INR_{spa}$ obtained from Stage I. The spatial INR was frozen at Stage II to prevent the interplay between the spatial and temporal INRs. For the multiresolution motion model, the three MBCs of increasing resolutions were learned progressively. In detail, the motion model starts with learning only the coarsest-resolution MBC per direction, and the other two MBCs of higher resolutions were progressively added into the learning process. When a finer-scale MBC was introduced into the learning, the coarse-scale MBCs were frozen without updating. The training on each scale used 100 epochs. The learning of Stage II used the projection-domain fidelity loss ($L_{prj}^b$):

$$L_{prj}^b = \frac{1}{N_{pixel}} \sum_t \sum_{j=1}^{N_{pixel}} \left\| \mathcal{P} \, INR_{spa}(\boldsymbol{x} + \sum_{i,k=1}^3 INR_{tem}^{i,k}(t) \times e_{i,k}(\boldsymbol{x})) - \boldsymbol{p}_t \right\|^2, \tag{11}$$

where $INR_{tem}^{i,k}(t)$ denotes the temporal INR output for the $i$th MBC along the $k$th Cartesian direction, at queried time frame $t$. Besides the fidelity loss, a regularization loss was applied to the MBCs to remove the ambiguity in the partial separation shown in Eq. 3. The loss enforced the orthonormal condition on $\{\boldsymbol{e}_i(\boldsymbol{x})\}_{i=1}^3$:

$$L_{MBC} = \sum_{i=1}^3 \left( \left\| \boldsymbol{e}_i \right\|^2 - 1 + \sum_{j=i+1}^3 \boldsymbol{e}_i \cdot \boldsymbol{e}_j \right), \tag{12}$$

where the inner product is defined in the Hilbert space of the MBCs. After initializing the finest-scale MBCs, the multi-resolution MBCs were unfrozen and trained for an additional 50 epochs for fine-tuning before entering the next Stage (III). The overall loss function of Stage II was defined as

$$L_{tot}^{II} = L_{prj}^b + \lambda_{TV} \, L_{TV} + \lambda_{MBC} \, L_{MBC}. \tag{13}$$

The value of $\lambda_{MBC}$ was set to 1 via empirical searching, and the same value was used throughout Stages II and III.

Stage III performs joint training in which all components in PMF-STINR were unfrozen, based on the same loss function as Stage II (Eq. 13). The joint training allows simultaneous image reconstruction and registration to improve the accuracy of the reference CBCT and the coherence of the solved intra-scan motion. The training in Stage III used a total of 2,000 epochs.

PMF-STINR was implemented via the PyTorch library and trained on a graphic processing unit (GPU) card (NVIDIA Tesla V100).We used the Adam optimizer for the three-staged training, and the learning rates were reset when the fidelity loss changed from the image domain to the projection domain in Stage I, and when entering the Stage II or III. The learning rates of the spatial INR were respectively $4\times10^{-4}$ and $1\times10^{-7}$ for the two steps of Stage I, and $1\times10^{-8}$ and $5\times10^{-9}$ for Stages II and III, respectively. The temporal INR and the B-spline motion model used the same learning rates, which were $2\times10^{-3}$ and $2\times10^{-4}$ for Stages II and III, respectively.

### 2.3 Evaluation datasets and schemes

We evaluated PMF-STINR using three datasets: a simulation study using the extended cardiac torso (XCAT) digital phantom (Segars *et al.*, 2010), a measurement study using the dynamic thorax CIRS 008A physical phantom (Computerized Imaging Reference Systems, Inc.), and a multi-institutional dataset of real patient scans with various imaging protocols/scanners.

### 2.3.1 XCAT simulation study

The simulated XCAT phantom covers the thoracic-abdominal portion of the anatomy, for a dimension of $128\times128\times64$ voxels and an isotropic $4\times4\times4$ mm$^3$ voxel size. A spherical lung tumor 30-mm in diameter was inserted into the lower lobe of the right lung for motion tracking and assessment. Six free-breathing scenarios (X1-X6) were simulated to assess the accuracy of PMF-STINR in reconstructing dynamic CBCTs to capture different motion variations. X1 simulates the simplest scenario of a quasi-periodic breathing cycle (~5 s) with small amplitude variations. The average range of the tumor center-of-mass motion was about 13 mm. X2 includes a rapid baseline shift (~5 mm) in the middle of the scan (~30 s). X3 combines both breathing amplitude variations and baseline shifts. The breathing period of X4 is gradually increasing, along with the motion amplitude. X5 simulates a slow breathing or a fast-rotation scan where the acquisition contains only a single breathing cycle. This scenario is deemed extremely challenging for motion-sorting-based reconstruction algorithms (e.g., 4D-CBCT), as the projection angles of the sorted phases will be limited to a small range. X6 combines variations of the breathing period, motion amplitude, and baseline shift.

Based on the dynamic XCAT volumes, cone-beam projections were simulated using the tomographic package ASTRA toolbox (van Aarle *et al.*, 2016). The total scan time was set to 60 s, covering a 360° scan angle (6°/s gantry rotation speed). A total of $N_p = 660$ projections were generated based on a frame rate of 11 fps to mimic a clinical 3D CBCT scan. Each projection contained $256\times192$ pixels for a pixel resolution of $1.6\times1.6$ mm$^2$.

The accuracy of the reconstructed dynamic CBCTs was quantitatively evaluated using the relative error (RE) and the structural similarity index (SSIM) (Wang *et al.*, 2004). RE was defined as

$$RE = \frac{1}{N_p} \sum_t \sqrt{\frac{\sum_{i=1}^{N_{voxel}} \left\| \hat{I}(x_i, t) - I^{gt}(x_i, t) \right\|^2}{\sum_{i=1}^{N_{voxel}} \left\| I^{gt}(x_i, t) \right\|^2}}, \tag{14}$$

where $I^{gt}$ denotes the 'ground-truth' CBCTs. The accuracy of solved motion was evaluated by contour-based metrics: tumor center-of-mass error (COME) and Dice similarity coefficient (DSC). To calculate COMEs and DSCs, the lung tumors in the reference-frame CBCTs were contoured and propagated to other dynamic CBCTs by the dynamic DVFs solved by PMF-STINR, and compared with the 'ground-truth'.

### 2.3.2 CIRS measurement study

A dynamic thorax phantom (CIRS) was employed in a clinical measurement study to assess PMF-STINR. For CIRS, a spherical tumor with electron density similar to that of the phantom body was placed in the left lung. The tumor motion is driven by an actuator and can be customized. Six motion trajectories (C1-C6) were used in the CIRS phantom study, including X1 (C1), X3 (C2), X4 (C3), and X5 (C4) from the XCAT study, and two additional irregular trajectories (C5 and C6). The peak-to-peak amplitudes in the superior-inferior (SI) direction of these trajectories ranged from 24 to 30 mm, and those in the anterior-posterior (AP) direction ranged from 0 to 10 mm. For each of the motion scenarios, cone-beam projections of the dynamic phantom were acquired on a Varian VitalBeam system (Varian Medical Systems, Inc.) in the half-fan mode, with the phantom center aligned to the imaging iso-center. Each scan took about 1 min for a 360° scan angle, acquiring 894-896 projections. Due to the off-axis tumor location and the half-fan scan, the tumor was only visible in about half of the projections. The projections were acquired under a 125 kVp energy, with mAs of 15 mA/20 ms. Each projection had 1024×768 pixels with a 0.388×0.388 mm$^2$ pixel resolution. The projections were down-sampled to 256×172 before the reconstruction, and the reconstructed dynamic CBCTs are of 200×200×68 voxels, with an isotropic voxel size of 3×3×3 mm$^3$ and a temporal resolution of 15 fps (the same as the X-ray acquisition frame rate). We would like to clarify here that the spatial and temporal resolutions of the dynamic CBCTs for PMF-STINR can be arbitrarily adjusted due to the continuous nature of the spatiotemporal INRs, and we set fixed spatial/temporal resolutions here for the purpose of image/motion evaluation.

The solved tumor motion was compared against the programmed trajectories. Tumor COMEs in the AP, left-right (LR), and SI directions were individually evaluated. The Pearson correlation coefficient between the solved and the 'ground-truth' tumor motion trajectories of the SI direction was computed. Due to the half-fan scan geometry, the accuracy was evaluated only on the projection frames where the tumor was in the field-of-view.

### 2.3.3 Patient study

**Table 1.** Summary of imaging parameters of the patient study.

| Patient ID[a] | Source[b] | Vender | Scan mode | Projection size[c] | Pixel size (mm²) | kVp/mA/mS | SAD[d] (mm)/ SDD (mm) | Reconstructed CBCT voxels | Voxel size (mm³) |
|---|---|---|---|---|---|---|---|---|---|
| P1 | MDACC | Varian | Full fan | 512×384×1983 | 0.776×0.776 | 120/80/25 | 1000/1500 | 200×200×100 | 2×2×2 |
| P2 | MDACC | Varian | Full fan | 512×384×2729 | 0.776×0.776 | 120/80/25 | 1000/1500 | 200×200×100 | 2×2×2 |
| P3 | MDACC | Varian | Full fan | 512×384×1653 | 0.776×0.776 | 120/80/25 | 1000/1500 | 200×200×100 | 2×2×2 |
| P4 | SPARE | Elekta | Full fan | 512×512×1015 | 0.8×0.8 | 125/20/20 | 1000/1536 | 200×200×100 | 2×2×2 |
| P4-S | SPARE | Elekta | Full fan | 512×512×340 | 0.8×0.8 | 125/20/20 | 1000/1536 | 200×200×100 | 2×2×2 |
| P5 | SPARE | Elekta | Full fan | 512×512×1005 | 0.8×0.8 | 125/20/20 | 1000/1536 | 200×200×100 | 2×2×2 |
| P5-S | SPARE | Elekta | Full fan | 512×512×340 | 0.8×0.8 | 125/20/20 | 1000/1536 | 200×200×100 | 2×2×2 |
| P6 | UTSW | Varian | Half fan | 1024×768×895 | 0.388×0.388 | 125/15/20 | 1000/1500 | 206×206×68 | 3×3×3 |
| P7 | SPARE | Varian | Half fan | 1006×750×2416 | 0.388×0.388 | 120/20/20 | 1000/1500 | 200×200×68 | 3×3×3 |
| P7-S | SPARE | Varian | Half fan | 1006×750×679 | 0.388×0.388 | 120/20/20 | 1000/1500 | 200×200×68 | 3×3×3 |
| P8 | SPARE | Varian | Half fan | 1006×750×2918 | 0.388×0.388 | 120/20/20 | 1000/1500 | 200×200×68 | 3×3×3 |
| P8-S | SPARE | Varian | Half fan | 1006×750×677 | 0.388×0.388 | 120/20/20 | 1000/1500 | 200×200×68 | 3×3×3 |

[a]Patient ID with a suffix 'S' indicates a sparsely-sampled set extracted from the fully-sampled one.

[b]MDACC: MD Anderson Cancer Center (Lu *et al.*, 2007). SPARE: SPARE Challenge (Shieh *et al.*, 2019). UTSW: University of Texas Southwestern Medical Center.

[c]width (in pixel number) × height (in pixel number) × $N_p$ (number of projections).

[d]SAD: source-to-axis distance. SDD: source-to-detector distance.

PMF-STINR was further evaluated on a multi-institutional patient dataset. Table I summarizes the imaging parameters of the patient study. The half-scan scans were reconstructed of a lower spatial resolution than the full-fan scans to accommodate the extended field-of-view. A total of 12 cone-beam projection sets from eight patients were curated from three sources. The MDACC data (P1-P3) were acquired by a Varian system in full-fan mode (Lu *et al.*, 2007). A slow-gantry acquisition scheme was used to cover a 200° scan angle. The scans took between 4.5-5.8 min, acquiring 1,653-2,729 projections. The SPARE data were taken from the SPARE Challenge (Shieh *et al.*, 2019) which evaluated 4D-CBCT reconstruction algorithms from sparse-view acquisitions in both full- and half-fan modes. The SPARE data contained 10 patients, and we selected four patients with clear anatomical structures that can be tracked in 2D projections for motion evaluation. The full- and half-fan scans were acquired from Elekta and Varian systems, respectively. For the SPARE data, each patient had two sets of projections: one was a fully-sampled scan, and the other was a down-sampled sparse-view set equivalent to a 1-min scan (patient ID ends with a suffix 'S'). As in Table I, the sparse-view sets had much fewer projections. The UTSW data (P6) were acquired by a Varian system in half-fan mode. The scan time was about 1 min, covering a 360° scan angle.

Since the patient study had no 'ground-truth' 3D motion for evaluation, accuracy of the solved intra-scan motion was evaluated in re-projected 2D planes. Specifically, for each reconstructed dynamic CBCT, we re-projected it into a 2D digitally reconstructed radiograph (DRR) to match the corresponding cone-beam projection's imaging geometry. Motion tracked on the cone-beam projections and the DRRs was then compared by two methods: (**1**). *Structure-based motion evaluation*. The structures being tracked include the diaphragms, lung nodules, and other lung features. For diaphragm tracking, the Amsterdam Shroud (AS) technique (Zijp *et al.*, 2004) was employed to highlight the motion-induced intensity variations on both the cone-beam projection sets and the re-projected DRR sets. From the view-consolidated Amsterdam Shroud image, motion of the diaphragm dome was extracted based on the sharp image contrast at the diaphragm boundary. The match between the diaphragm motion tracked on the original cone-beam projections and that tracked on the re-projected DRRs was evaluated by the Pearson correlation coefficient and the localization error. Since the diaphragm of P1 moved in-and-out of the field-of-view during the scan, instead of the diaphragm, a high-density lung nodule was tracked with a similar approach as diaphragm tracking. In addition, diaphragm of P3 was barely visible in the projections, so the respiratory motion trajectory was extracted from the AS images using a high-contrast lung feature. **(2).** *Feature point-based motion evaluation.* The second method automatically tracked corresponding feature points from both cone-beam projections and DRRs for motion comparison, without using the AS images (Park *et al.*, 2017; Wei *et al.*, 2020). Specifically, it first automatically and independently extracted image feature points from cone-beam projections based on local intensity variations, with the selected points typically at the boundaries of anatomic features. Next, corresponding feature points in these projections were identified and selected, and their motion trajectories across multiple frames were tracked, using the M-estimator sample consensus algorithm (Torr and Zisserman, 2000). Finally, the corresponding feature points in the DRRs were localized and tracked by a correlation coefficient-based searching algorithm for comparison. The feature point motion difference (localization error, LE) between the cone-beam projections and DRRs was evaluated by (Wei *et al.*, 2020):

$$LE = \sqrt{\frac{1}{N_p}\sum_p \frac{1}{M_p}\sum_q \left(z_{pq}^{cb} - z_p^{DRR}\right)^2}, \tag{15}$$

where $M_p$ denotes the number of feature points in the $p^{th}$ cone-beam projection, $z_{pq}^{cb}$ denotes the tracked location of the $q^{th}$ feature point in the $p^{th}$ cone-beam projection, and $z_p^{DRR}$ is the tracked location of the corresponding point in the $p^{th}$ DRR. Due to the limitations of tracking in 2D planes with rotating projection angles, we only calculated LE along the SI direction, which is the dominant direction of respiratory motion.

### 2.3.4 Comparison studies

Due to the challenge of dynamic CBCT reconstruction, there are very few available dynamic CBCT reconstruction techniques for comparison. And since in our preliminary study the original STINR framework has proved more accurate than the INR method with polynomial-based DVF fitting (INR-Poly) and the deformation-driven PCA method, we did not include them further in this comparison. Instead, in this study we focused the comparison between PMF-STINR and the previously-published STINR framework (Zhang *et al.*, 2023) to evaluate the benefit of the prior model-free approach of PMF-STINR. We performed this comparison study on the XCAT dataset with known 'ground-truth'. Two variants of the original STINR framework was compared. The first variant uses a XCAT-simulated 4D-CT of the same patient (motion: sinusoidal curve of 5 s cycle) for PCA motion modeling, which matches with the configuration of the published STINR study. The corresponding method, STINR$_{\text{PCA-4DCT}}$, represents an ideal scenario where a high-quality prior 4D-CT is available for motion modeling, and there is no inter-fractional deformation or anatomy change between the 4D-CT and the dynamic CBCT acquisitions. The second framework does not have access to the 4D-CT motion model. Instead, its motion model was derived from the 4D-CBCT set reconstructed using the dynamic cone-beam projections of each motion scenario. The second variant was named STINR$_{\text{PCA-4DCBCT}}$. Compared with the motion model of PMF-STINR, the PCA-based motion model of STINR$_{\text{PCA-4DCBCT}}$ is not learned/optimized but generated prior to the reconstruction using ad-hoc reconstructed 4D-CBCTs. Specifically, prior to the STINR$_{\text{PCA-4DCBCT}}$ reconstruction, the cone-beam projections of each motion scenario were first sorted and binned into 10 respiratory phases, except for scenario X5. For X5, we sorted the projections into 5 phases, as the single-cycle scenario yielded substantial limited-angle artifacts if 10 phases were used, which led to highly inaccurate PCA motion modeling. We reconstructed 4D-CBCTs from the phase-sorted projections using the FDK algorithm. Deformable image registrations between the end-exhale phase and the other phases were performed to obtain the inter-phase DVFs for each motion scenario, using Elastix (Klein *et al.*, 2010). Finally, we derived principal motion components from the inter-phase DVFs using PCA for each motion scenario, and fed them as a known motion model into the STINR$_{\text{PCA-4DCBCT}}$ reconstruction. For both STINR$_{\text{PCA-4DCBCT}}$ and STINR$_{PCA\text{-}4DCT}$, we added multi-resolution hash encoding and spatial TV regularization into the original STINR framework, for fair comparison with PMF-STINR.

We compared the accuracy of the reconstructed dynamic CBCTs and the solved intra-scan motion between STINR$_{PCA\text{-}4DCBCT}$, STINR$_{PCA\text{-}4DCT}$, and PMF-STINR. Wilcoxon signed-rank tests were performed to evaluate the statistical significance of the result differences between the three methods.

## 3. Results

### 3.1 The XCAT study results

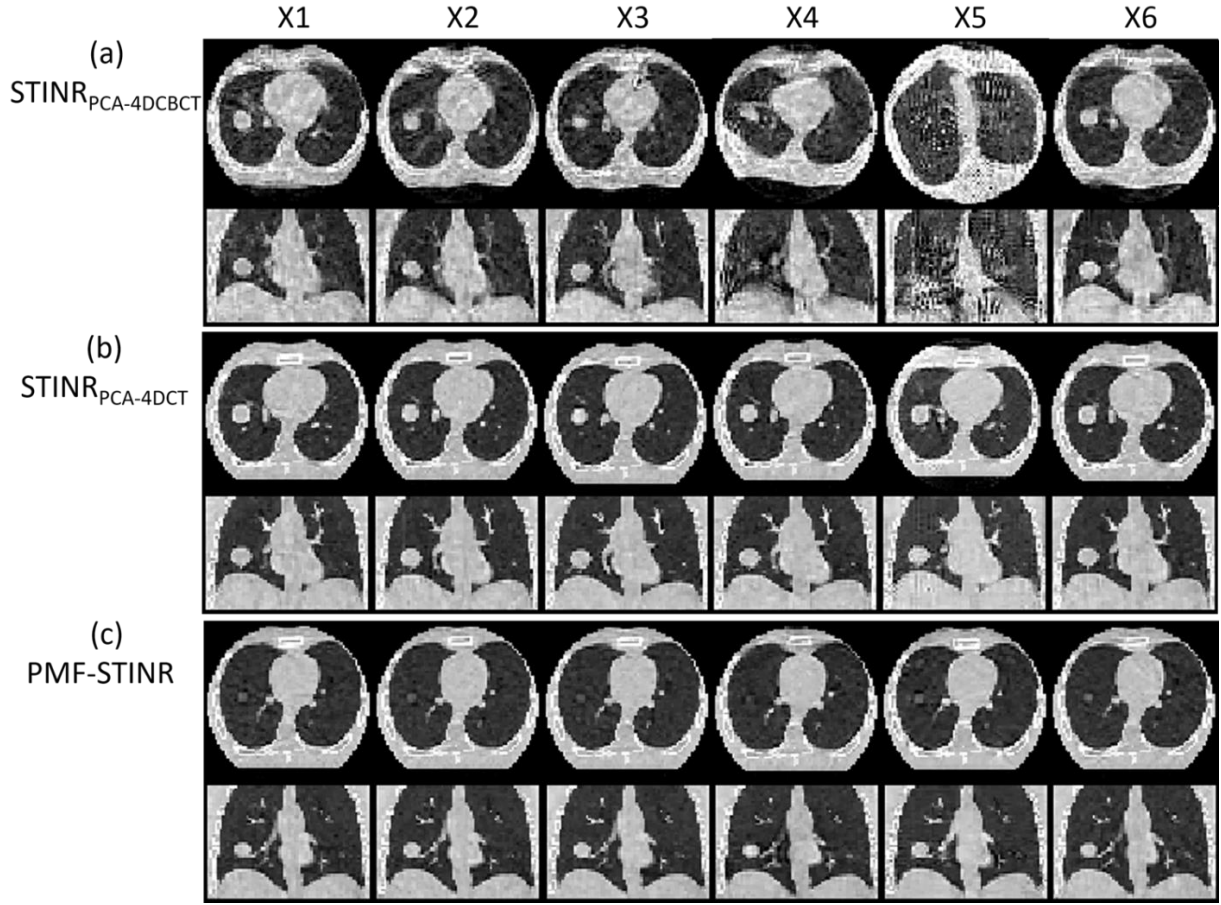

**Fig. 4** Comparison of reference-frame CBCTs reconstructed by (a) STINR$_{PCA\text{-}4DCBCT}$, (b) STINR$_{PCA\text{-}4DCT}$, and (c) PMF-STINR for the six XCAT motion scenarios (X1-X6) in the axial and coronal views. The circular boundary in the axial view reflects the field-of-view from the clinically-realistic projection size simulation in full-fan mode (Sec. 2.3.1). The display window for all images is [0, 0.4] mm$^{-1}$.

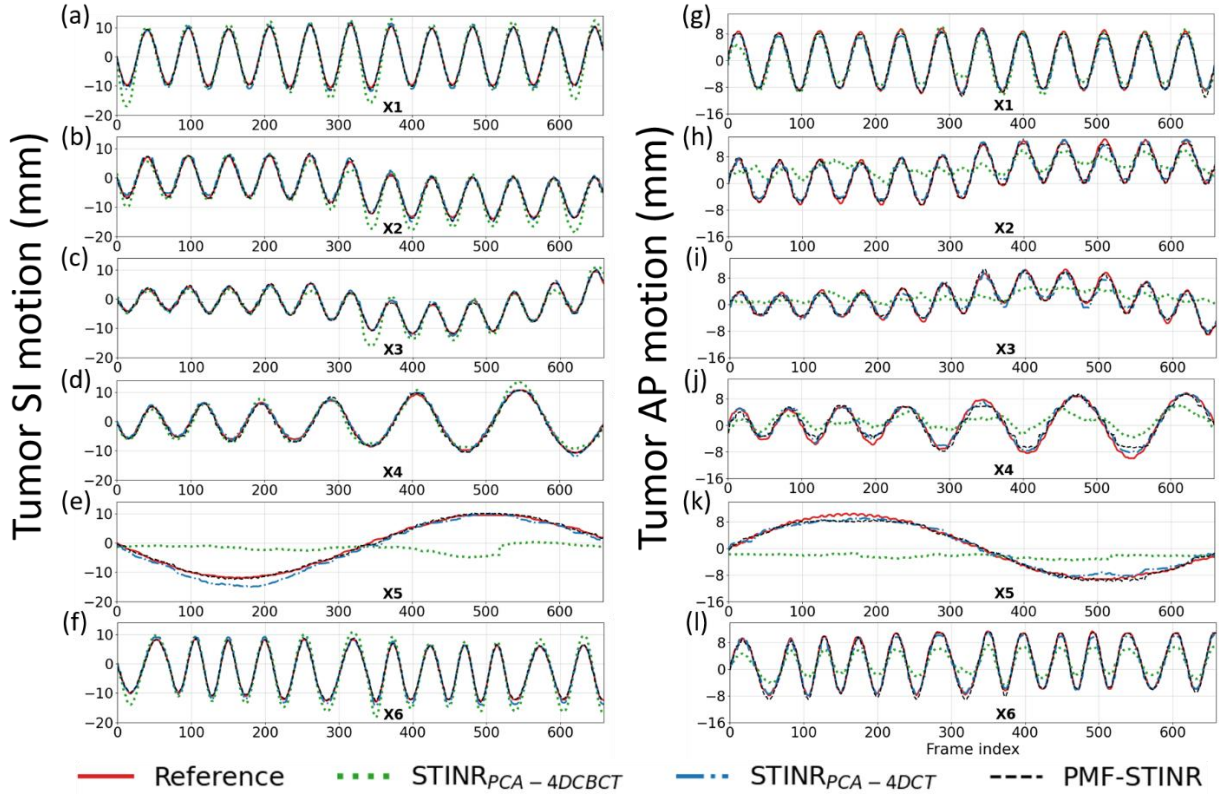

**Fig. 5** Comparison of tumor SI (a-f) and AP (g-l) trajectories of the six motion scenarios of the XCAT study (X1-X6), between STINR$_{PCA-4DCBCT}$, STINR$_{PCA-4DCT}$, PMF-STINR, and the 'ground-truth' reference.

**Table 2** Accuracy of the reconstructed dynamic CBCTs of the XCAT study, in terms of the relative error and the structural similarity index (SSIM). The results are presented as mean ± S.D.. The arrows are pointing to the direction of higher accuracy.

| Motion scenario | Relative error ↓ | | | SSIM ↑ | | |
|---|---|---|---|---|---|---|
| | STINR$_{PCA-4DCBCT}$ | STINR$_{PCA-4DCT}$ | PMF-STINR | STINR$_{PCA-4DCBCT}$ | STINR$_{PCA-4DCT}$ | PMF-STINR |
| X1 | 0.244±0.021 | 0.160±0.009 | **0.149±0.008** | 0.890±0.006 | 0.893±0.038 | **0.984±0.002** |
| X2 | 0.264±0.034 | 0.154±0.017 | **0.143±0.016** | 0.905±0.010 | 0.905±0.025 | **0.985±0.004** |
| X3 | 0.254±0.040 | 0.146±0.017 | **0.137±0.016** | 0.912±0.012 | 0.925±0.023 | **0.986±0.004** |
| X4 | 0.279±0.029 | **0.158±0.013** | 0.168±0.020 | 0.865±0.007 | 0.917±0.015 | **0.979±0.006** |
| X5 | 0.365±0.034 | 0.192±0.009 | **0.170±0.010** | 0.790±0.011 | 0.860±0.052 | **0.978±0.003** |
| X6 | 0.251±0.021 | 0.158±0.009 | **0.149±0.010** | 0.892±0.006 | 0.902±0.023 | **0.984±0.003** |

**Table 3** Lung tumor tracking accuracy of the XCAT study. The results are presented in terms of mean ± S.D.. The arrows are pointing to the direction of higher accuracy.

| Motion scenario | COME (mm) ↓ | | | DSC ↑ | | |
|---|---|---|---|---|---|---|
| | STINR$_{PCA-4DCBCT}$ | STINR$_{PCA-4DCT}$ | PMF-STINR | STINR$_{PCA-4DCBCT}$ | STINR$_{PCA-4DCT}$ | PMF-STINR |
| X1 | 2.2±1.5 | 1.1±0.5 | **0.8±0.4** | 0.855±0.051 | 0.906±0.024 | **0.910±0.025** |
| X2 | 4.2±2.6 | 1.1±0.5 | **0.8±0.4** | 0.796±0.094 | 0.913±0.025 | **0.925±0.030** |
| X3 | 3.5±2.4 | 1.0±0.5 | **0.8±0.3** | 0.823±0.095 | 0.915±0.015 | **0.927±0.031** |
| X4 | 3.9±1.8 | **1.0±0.5** | 1.3±0.6 | 0.783±0.067 | **0.913±0.023** | 0.913±0.026 |
| X5 | 10.1±4.4 | 1.6±0.8 | **1.0±0.5** | 0.430±0.121 | 0.892±0.035 | **0.899±0.023** |
| X6 | 3.4±1.8 | 1.0±0.4 | **0.9±0.4** | 0.821±0.054 | **0.914±0.023** | 0.905±0.022 |

Figure 4 compares the reconstructed reference-frame CBCTs between $STINR_{PCA-4DCBCT}$, $STINR_{PCA-4DCT}$, and PMF-STINR for the XCAT study. Correspondingly, Fig. 5 compares the tracked tumor motion between $STINR_{PCA-4DCBCT}$, $STINR_{PCA-4DCT}$, and PMF-STINR, with the 'ground-truth' trajectories as reference. Table 2 shows the accuracy of reconstructed dynamic CBCTs of different methods, and Table 3 shows the corresponding tumor motion tracking accuracy. For Tables 2 and 3, the Wilcoxon signed-rank tests yielded $p$-values of $< 10^{-3}$ between $STINR_{PCA-4DCBCT}$ and $STINR_{PCA-4DCBCT}$ for all metrics and motion scenarios. Similarly, the $p$-values are $< 10^{-3}$ between $STINR_{PCA-4DCT}$ and $STINR_{PCA-4DCBCT}$ for all metrics and motion scenarios, except for DSC of scenario X4 ($p = 0.663$).

In Fig. 4, PMF-STINR presented consistently better image quality than $STINR_{PCA-4DCBCT}$ in terms of contrast, shape, and intensity distributions of the anatomy, across all six motion scenarios. The reference-frame CBCT image quality of $STINR_{PCA-4DCBCT}$ is considerably worse, especially for scenario X5. Compared with PMF-STINR for which the motion model was learned and optimized on the fly during the reconstruction, $STINR_{PCA-4DCBCT}$ had to rely on the same PCA motion model derived from the ad-hoc reconstructed 4D-CBCT throughout the image reconstruction. Imaging artifacts of the 4D-CBCT, including those from under-sampling, intra-phase residual motion, and motion irregularity, reduced the accuracy of the resulting PCA motion model. The inaccurate PCA motion model in turn negatively impacted the reference-frame CBCT reconstruction and intra-scan DVF optimization for $STINR_{PCA-4DCBCT}$. For scenario X5 where there is only one breathing cycle (Fig. 5), the 4D phase sorting led to limited-angle reconstruction for each phase image, and resulted in a highly inaccurate PCA motion model. The erroneous PCA motion model significantly impacted the reference-frame CBCT quality for $STINR_{PCA-4DCBCT}$. In comparison, using the PCA motion model derived from a simulated artifact-free 4D-CT, $STINR_{PCA-4DCT}$ generated better results than $STINR_{PCA-4DCBCT}$, with accuracy level on par with that reported in the original STINR study (Zhang *et al.*, 2023). However, the high-quality, artifacts-free 4D-CT of XCAT is difficult to realize in real clinical cases. And for the XCAT study, there was no inter-scan deformation or anatomy change, which favorably biased the $STINR_{PCA-4DCT}$ results. Even so, PMF-STINR provided overall more accurate reconstruction than $STINR_{PCA-4DCT}$ (Tables 2 and 3). Compared with $STINR_{PCA-4DCT}$, PMF-STINR used cone-beam projections to directly learn and optimize the motion model, which can better fit the onboard data and avoid uncertainties from the 4D-CT deformable registration used to derive the PCA model. It is also interesting to note that the reference-frame CBCT of $STINR_{PCA-4DCT}$ and PMF-STINR are corresponding to different motion states (Fig. 4). For $STINR_{PCA-4DCT}$, the reference frame is expected to be close to the end-exhale phase (Sec. 2.3.4), on which the PCA motion model was built. However, for PMF-STINR, the reference frame is not pre-fixed to a specific phase and the optimization framework will jointly determine the motion model and the reference frame to best match the dynamic projections. Figure 6 plots two examples of the dynamic CBCTs reconstructed by PMF-STINR (scenarios X1 and X2), with the 'ground-truth' reference dynamic CBCTs for comparison. It can be observed that PMF-STINR successfully reconstructed the dynamic CBCTs to match well with the 'ground-truth' in anatomy and motion.

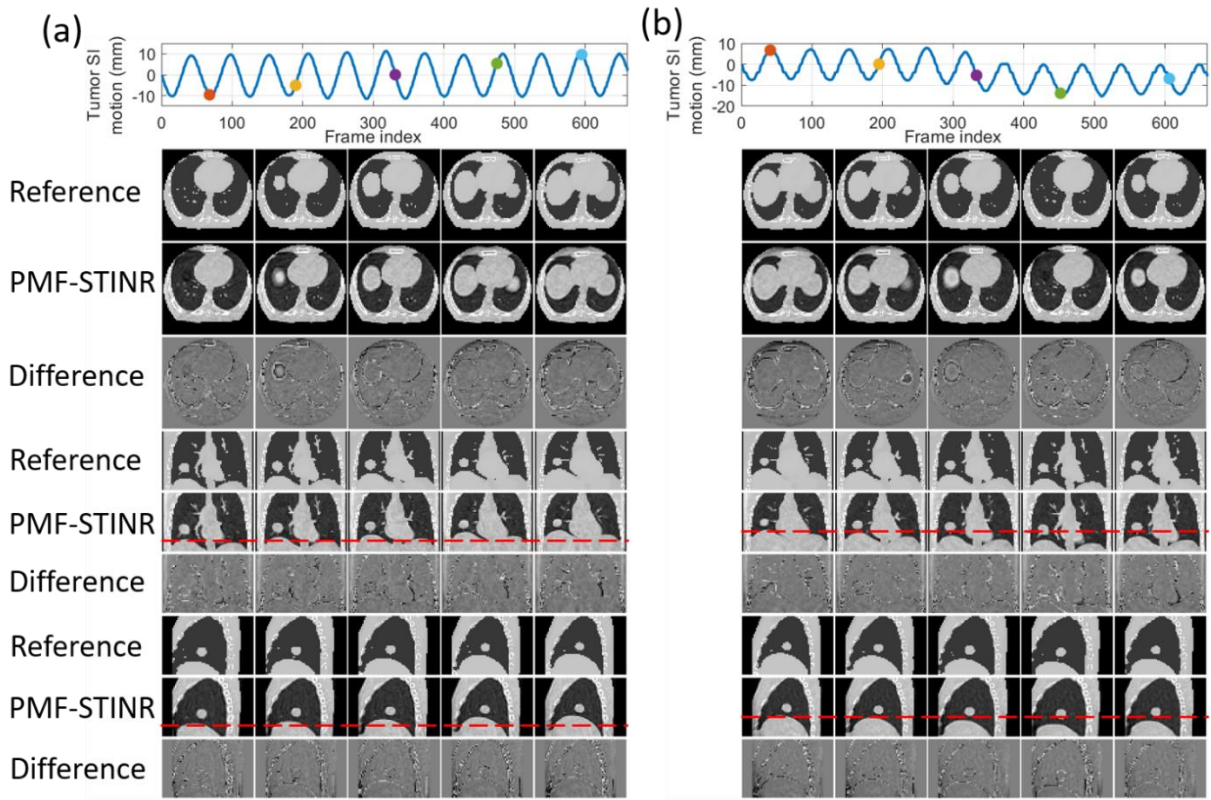

**Fig. 6** Examples of dynamic CBCTs reconstructed by PMF-STINR for the XCAT study: (a). scenario X1; and (b) scenario X2. First row shows the corresponding motion curves along the SI direction, with the dots indicating the motion states selected for plotting. In the following rows, PMF-STINR CBCTs of the selected motion states were compared against the 'ground-truth' reference dynamic CBCTs, with the difference images calculated. The display window for the CBCT images is [0, 0.4] mm$^{-1}$, while that for the difference image is [-0.2, 0.2] mm$^{-1}$.

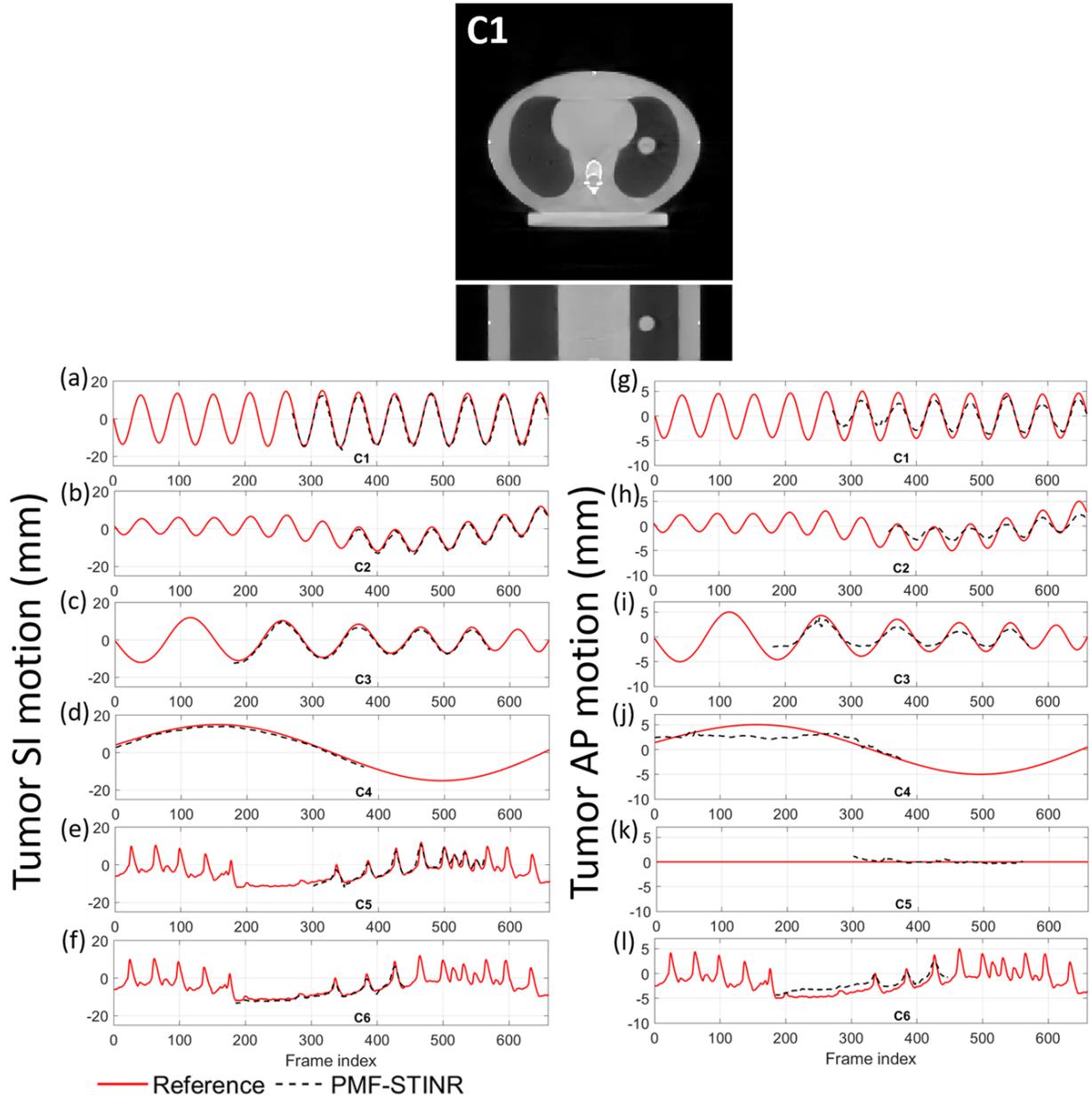

**Fig. 7** (Upper panel) Example reference-frame CBCT images reconstructed by PMF-STINR for the CIRS study (scenario C1). (Lower panels) Comparison between the programmed motion curves (reference) and the solved curves by PMF-STINR for scenarios C1-C6. For PMF-STINR, the tumor trajectories were only available for frames where the tumor was in the field-of-view.

Figure 7 presents the reconstructed reference-frame CBCT images of PMF-STINR for scenario C1, and the comparison between the programmed (reference) motion curves and the solved curves by PMF-STINR of scenarios C1-C6 for the CIRS study. Due to variations of the scan start time relative to the phantom motion trajectories, the tracked portion of each motion trajectory varies. Since the moving portion of the CIRS phantom is limited to the tumor, when the tumor moved outside the field-of-view, there were no moving features available in the projections to reconstruct the motion. Thus for those projections with the tumor outside the field-of-view, motion was not evaluated. For the projections where tumors are contained, PMF-STINR accurately captured the intra-scan motion of the tumor for both regular and irregular motion scenarios. Table 4 presents the tumor center-of-mass localization errors in the LR, AP, and SI directions and the Pearson correlation coefficients between the solved and the 'ground-truth' SI trajectories. A tracking error within or around 1 mm was achieved for all three Cartesian directions. Figure 8 plots one example of the dynamic CBCTs reconstructed by PMF-STINR for the CIRS study (scenarios C3). The dynamic motion of the tumor matches well with the programmed curves.

**Table 4** Lung tumor localization accuracy by PMF-STINR for the dynamic thorax CIRS phantom study. The results are presented in terms of mean ± S.D.. The arrows are pointing to the direction of higher accuracy.

| Motion scenario | Pearson correlation coefficient (SI trajectory) ↑ | Tumor localization error ↓ | | |
| --- | --- | --- | --- | --- |
| | | LR (mm) | AP (mm) | SI (mm) |
| C1 | 0.997 | 0.4±0.4 | 1.1±0.7 | 1.2±0.7 |
| C2 | 0.999 | 0.4±0.2 | 1.0±0.7 | 1.0±0.3 |
| C3 | 0.999 | 0.5±0.6 | 1.0±0.6 | 1.0±0.4 |
| C4 | 0.999 | 0.3±0.3 | 1.1±0.9 | 1.0±0.2 |
| C5 | 0.959 | 0.4±0.4 | 0.2±0.2 | 1.1±1.1 |
| C6 | 0.979 | 0.2±0.3 | 1.1±0.4 | 0.9±0.6 |

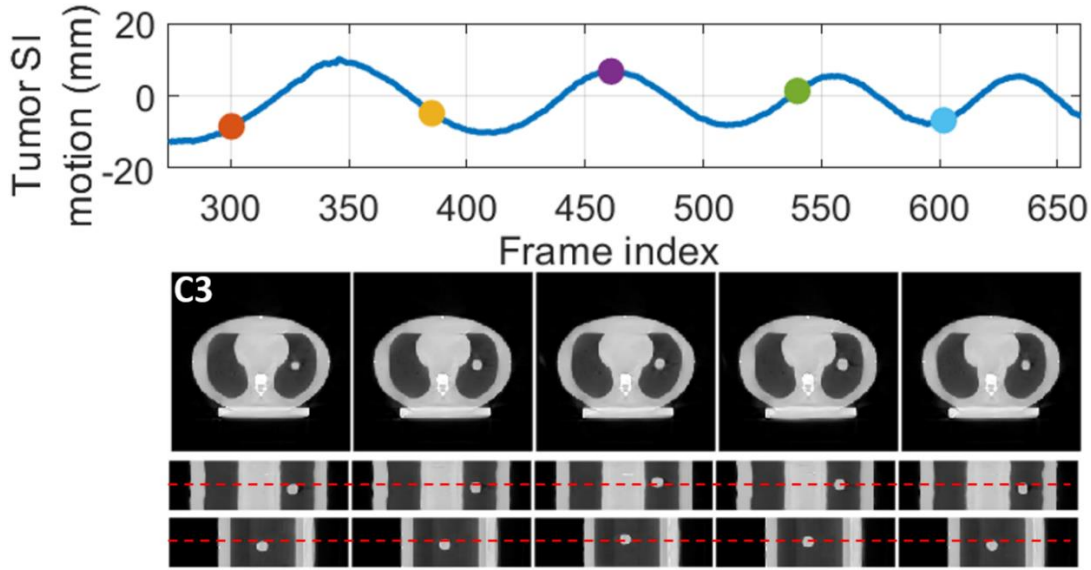

**Fig. 8** Examples of dynamic CBCTs reconstructed by PMF-STINR for the CIRS study: scenario C3. First row shows the corresponding motion curve along the SI direction, with the dots indicating the motion states selected for plotting. In the following rows, PMF-STINR CBCTs of the selected motion states were shown in three views. The display window for the CBCT images is [0, 0.022] mm⁻¹.

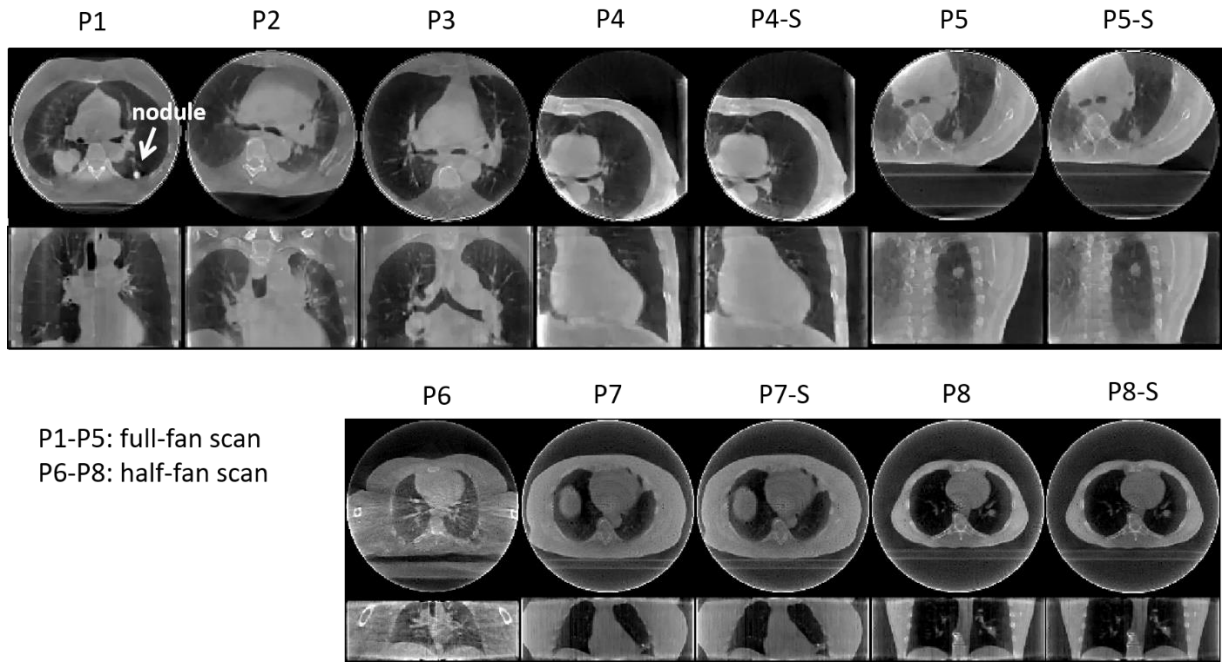

**Fig. 9** Reference-frame CBCTs reconstructed by PMF-STINR for the patient study. The upper and lower rows show the reconstructions from full- and half-fan scans, respectively. The display windows for the CBCT images range between [0, 0.026] mm$^{-1}$ and [0, 0.039] mm$^{-1}$ to optimize the contrast.

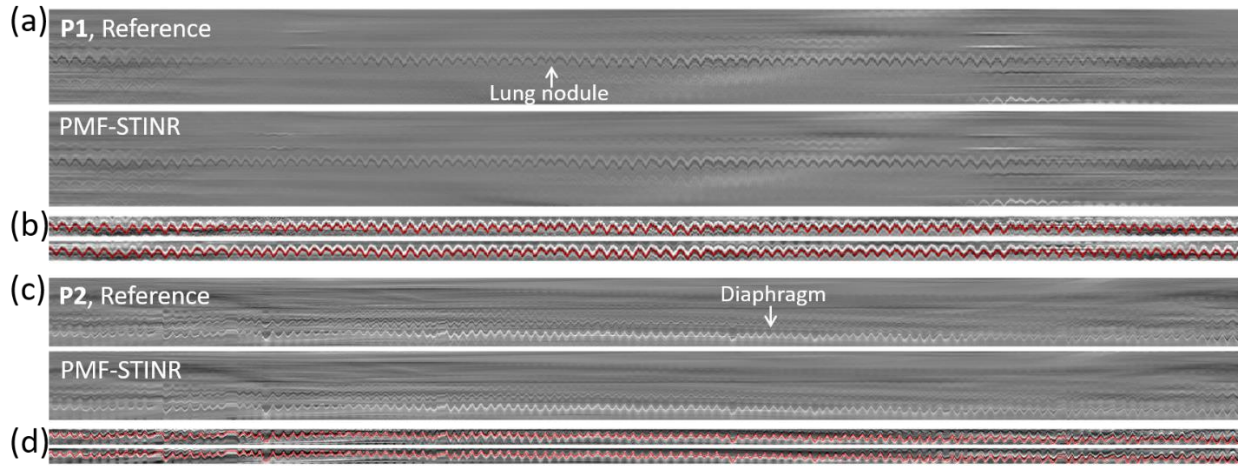

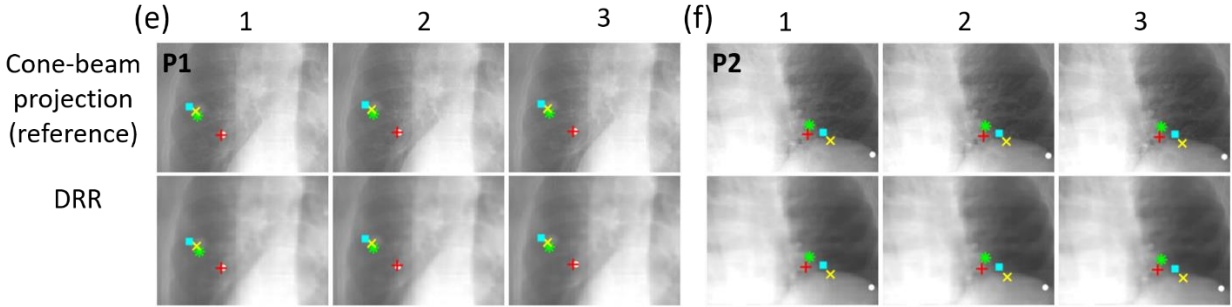

**Fig. 10 (a-d)** Examples illustrating the reference and the PMF-STINR-solved motion trajectories, respectively extracted from the cone-beam projections and the DRRs via the Amsterdam Shroud (AS) method. Figs. 10 a & b show the AS-tracked results for P1, based on a lung nodule. Figs. 10 c & d show the AS-tracked results for P2, based on the diaphragm. Figs. 10 a & c show the AS images derived from the projections and DRRs. Figs. 10 b & d show the vertically cropped, z-score normalized regions containing the tracked objects, on which the motion trajectories (red lines) were extracted for quantitative evaluation. **(e-f)** Examples illustrating the feature points automatically extracted and tracked from the consecutively acquired cone-beam projections and the corresponding DRRs via the feature point-tracking method for localization comparison.

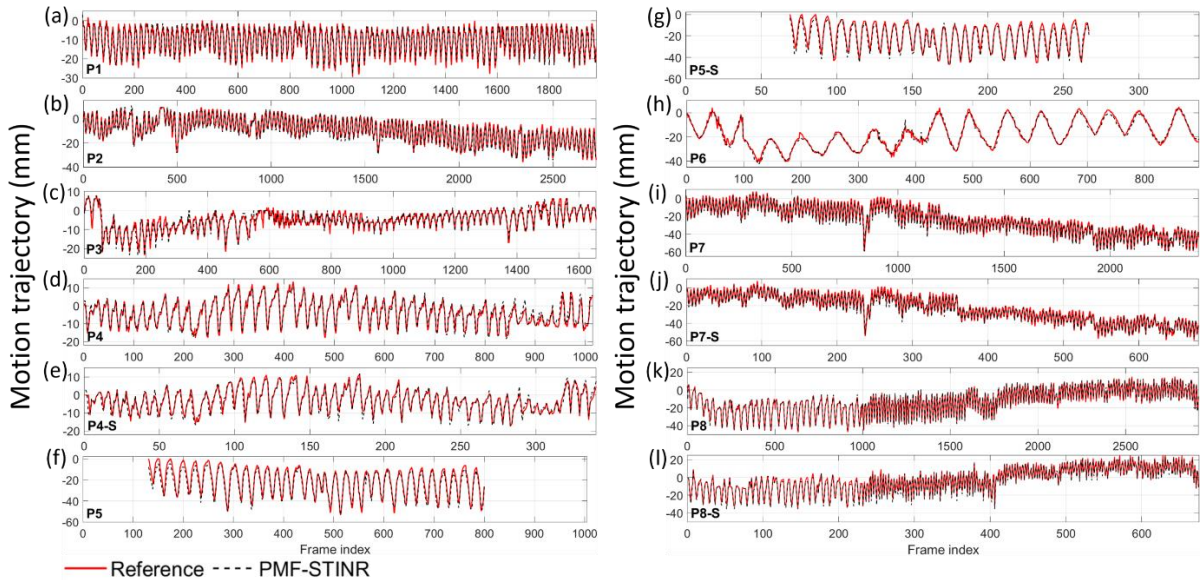

**Fig. 11** Comparison between tracked and reference SI trajectories of P1-P8 for the patient study using the Amsterdam Shroud image-based method, between the PMF-STINR curves (extracted from the re-projected DRRs from reconstructed dynamic CBCTs) and the reference curves (extracted from the cone-beam projections).

Figure 9 presents the reference-frame CBCTs reconstructed for the patient study. The reference-frame CBCTs of the fully- and sparsely-sampled acquisitions (P4, P5, P7, and P8) are comparable in image quality, showing that PMF-STINR allows dynamic CBCT reconstruction from sparsely-sampled 3D CBCT scans (~340 total projections, Table 1). The high-density lung nodule of P1 used for motion tracking and evaluation was highlighted (Sec. 2.3.3). Figure 10 shows example motion curves extracted by the AS image-based method, and example feature points identified by the automatic feature point tracking method. Figure 11 compares between the tracked and the reference SI trajectories of various lung anatomies (e.g.

lung nodule, diaphragm) by the AS image-based method. The tracked curves by PMF-STINR match well with the reference curves directly extracted from the cone-beam projections. Various motion irregularities, including amplitude variations, frequency changes, and baseline shifts and drifts, were accurately captured. Part of the gradual baseline drifts observed in the curves is due to the rotating angle geometry and the off-iso locations of the tracked anatomy. For P5 and P5-S, the tracked anatomy (diaphragm) can only be seen in a subset of the projections. Although PMF-STINR can still solve its motion from other motion features and moving parts in the diaphragm-occluded projections, we cannot resolve the corresponding motion from the cone-beam projections as reference for comparison. Thus only the diaphragm-visible section of the trajectory was evaluated (Fig. 11).

Table 5 shows the accuracy of the solved motion by PMF-STINR for the patient study, using both the AS image-based tracking and the feature point tracking methods. Overall, PMF-STINR achieved accurate structure and feature point localization in the dynamic sequences, as evaluated by the counterpart reference signals extracted from the corresponding cone-beam projections. The localization errors presented here were calculated in projected 2D planes, and included the magnification factor (~1.5) due to the imaging trajectory. On average a sub-millimeter accuracy was achieved by PMF-STINR-solved motion, after accounting for the factor. Figure 12 plots two examples of the dynamic CBCTs reconstructed by PMF-STINR for the patient study (P1 and P5). The dynamics of the lung nodule (P1) and the lung tumor (P5) are well captured.

**Table 5** Accuracy of the solved intra-scan motion by PMF-STINR for the patient study. The results are presented in terms of mean ± S.D.. The arrows are pointing to the direction of higher accuracy.

| Patient ID | Pearson correlation coefficient (SI trajectory)↑ | Localization error (mm) ↓ | |
|:---:|:---:|:---:|:---:|
| | | AS image-based tracking | Feature point tracking |
| P1 | 0.968 | 1.1±1.1 | 0.5±0.7 |
| P2 | 0.987 | 1.1±1.2 | 1.1±1.1 |
| P3 | 0.943 | 1.1±1.3 | 0.9±0.9 |
| P4 | 0.976 | 1.1±1.0 | 1.7±1.1 |
| P4-S | 0.955 | 1.3±1.5 | 1.6±1.3 |
| P5 | 0.978 | 2.7±2.2 | 1.2±1.3 |
| P5-S | 0.978 | 2.3±2.1 | 1.1±1.6 |
| P6 | 0.987 | 1.4±1.2 | 0.7±0.8 |
| P7 | 0.992 | 1.7±1.4 | 1.9±1.6 |
| P7-S | 0.992 | 1.7±1.5 | 1.8±1.6 |
| P8 | 0.992 | 1.4±1.1 | 1.9±1.4 |
| P8-S | 0.990 | 1.8±1.6 | 2.1±1.8 |

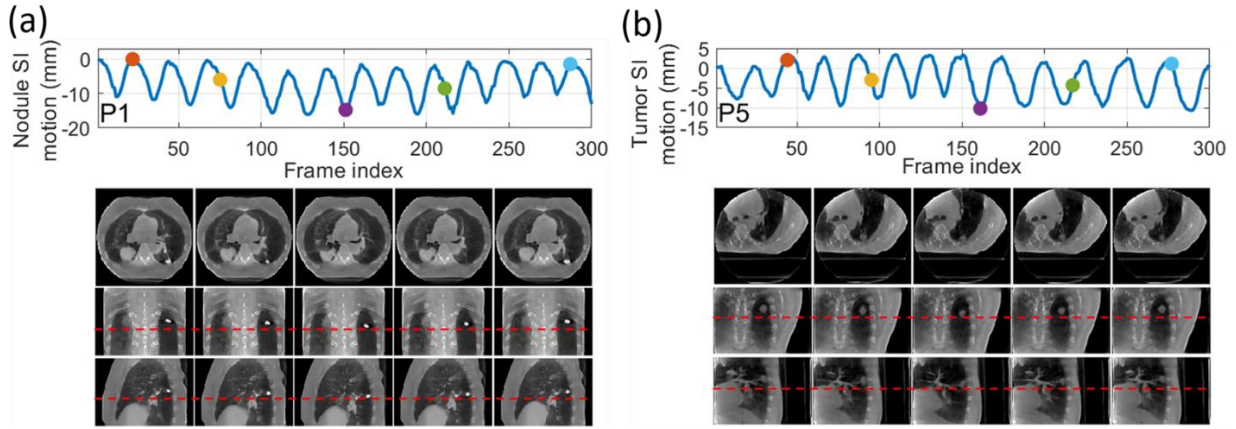

**Fig. 12** Examples of dynamic CBCTs reconstructed by PMF-STINR for the patient study: (a). P1; and (b) P5. First row shows the corresponding motion curves along the SI direction, with the dots indicating the motion states selected for plotting. In the following rows, PMF-STINR CBCTs of the selected motion states were shown in three views. The display windows for the CBCT images are [0, 0.031] mm$^{-1}$ for P1, and [0.005, 0.028] mm$^{-1}$ for P5. Due to the space limit, only partial trajectories are shown. For movies showing the full dynamic sequences, please refer to the supplementary file.

## 5. Discussion

In this study, we proposed PMF-STINR (Fig. 1), an INR-based framework to address the dynamic CBCT reconstruction challenge. Without relying on any prior anatomical or motion model, PMF-STINR reconstructs dynamic CBCTs and solves the intra-scan motion simultaneously from the cone-beam projections via a 'one-shot' learning scheme. PMF-STINR decouples the ill-posed spatiotemporal inverse problem of dynamic CBCT reconstruction into learning a spatial INR module (Fig. 2a), a temporal INR module (Fig. 2b), and a trainable B-spline-based, data-driven motion model, and develops a progressive learning strategy (Fig. 3) that is able to reconstruct dynamic CBCTs from 3D CBCT scans with ~300 projections. Compared with the prior STINR framework, PMF-STINR achieves substantially higher reconstruction and motion tracking accuracy by the motion model learned and optimized on the fly, and does not require any prior motion sorting and binning (Figs. 4 and 5, Tables 2 and 3). The use of the hash encoding, compared to the Fourier frequency encoding, enables faster computation and allows lightweight MLPs to be used for the INRs. The comprehensive evaluation of PMF-STINR on the XCAT (Figs. 4-6, Tables 2 and 3), the CIRS (Figs. 7-8, Table 4), and the patient datasets (Figs. 9-12, Table 5) demonstrated its robustness to varying anatomical, motion, and imaging variations, a substantial advantage over previous deep learning-based models which require extensive training and are susceptible to out-of-distribution shifts. The robustness and adaptability of PMF-STINR indicate a high clinical translation potential.

Whereas the spatial and temporal INRs were lightweight and learning efficient, currently PMF-STINR took about 4 hours to reconstruct a sequence of 2000-frame dynamic CBCTs on a Nvidia V100 card. A major speed bottleneck was the cone-beam projector from the ASTRA toolbox (van Aarle *et al.*, 2016), which was optimized for projecting multiple DRRs from the same CBCT in parallel. However, PMF-STINR only requires a single DRR at a gantry angle for a dynamic CBCT, and thus the DRRs of different dynamic CBCTs are sequentially projected in the current framework. To reduce the reconstruction time, the high throughput of a GPU can be leveraged by parallelizing a cone-beam projector to generate DRRs of different CBCTs simultaneously.

In the present study, PMF-STINR was assessed for solving the respiratory motion of the thoracic-abdominal region. The proposed framework, in theory, can handle other types of anatomical motions, such as the cardiac beats and peristalsis. However, for motion fields involving multiple modes of anatomical motions with disparate temporal and spatial scales (e.g., respiratory and cardiac motions have distinct periods of about 3-5 s and 0.8-1 s, respectively), the PMF-STINR framework may warrant additional modifications to effectively describe the multi-mode motion field. It is possible to describe a multi-mode and multi-scale DVF by increasing the number of motion basis components along each direction, or further introducing region-focused motion basis components. The current motion model used a uniform B-spline grid to cover the entire reconstructed volume. For improvement, a non-uniform grid can be used wherein the grid density will automatically adapt to the involved anatomical locations. Further investigations and developments of PMF-STINR are warranted to evaluate and potentially improve its versatility and adaptability for different motion types and scenarios.

Currently, PMF-STINR was designed for the retrospective reconstruction of dynamic CBCTs, preventing it from solving time-resolved motion in real-time. However, with additional modifications to the framework, PMF-STINR may leverage the obtained knowledge of the reference-frame CBCT and the learned motion model to achieve real-time imaging and tumor localization. Currently, the temporal INR takes a temporal frame index as input and outputs the temporal coefficients of the motion basis components at the queried time point. To adapt the framework for real-time imaging, the input can be replaced by motion-related image features extracted from a cone-beam projection by a deep learning network. This modification may allow the temporal INR to learn the mapping from the extracted imaging features to the coefficients of motion basis components. After the network training, the temporal INR may directly use real-time-acquired imaging features to infer the motion basis component coefficients and compose real-time DVFs to represent instantaneous motion. Such modifications towards real-time imaging are currently under investigation with future reports anticipated.

## 6. Conclusion

We developed a prior model-free, spatiotemporal implicit neural representation method in this study to reconstruct dynamic CBCTs. The technique can reconstruct dynamic images from conventional or sparsely-sampled 3D-CBCTs in a 'one-shot' fashion, without using any prior anatomical/motion model or requiring any motion sorting/binning. The method has demonstrated accuracy and robustness through comprehensive evaluations using digital phantom, physical phantom, and real patient studies. The dynamic CBCTs offer richer motion information than current state-of-the-art 4D-CBCTs, and can capture various motion irregularities including motion amplitude variations, frequency changes, and baseline shifts and drifts to inform better motion management strategies. The simultaneously solved intra-phase motion fields in addition to the CBCTs can be applied for structure propagation, dose accumulation, and adaptive radiotherapy.

## Acknowledgments

The study was supported by funding from the National Institutes of Health (R01 CA240808, R01 CA258987, R01 CA280135), and from Varian Medical Systems. We would like to thank Dr. Ran Wei from the Cancer Hospital, Chinese Academy of Medical Sciences for sharing the MATLAB codes to perform feature point tracking. We would also like to thank Dr. Paul Segars at Duke University for providing the XCAT phantom for our study, and Dr. Park Yang at UT Southwestern Medical Center for helping with configuring the CIRS phantom for measurements.

## Ethical statement

The MDACC dataset used in this study was retrospectively collected from an IRB-approved study at MD Anderson Cancer Center in 2007. The UTSW dataset was retrospectively collected from an approved study at UT Southwestern Medical Center on August 31, 2023, under an umbrella IRB protocol 082013-008 (Improving radiation treatment quality and safety by retrospective data analysis). This is a retrospective analysis study and not a clinical trial. No clinical trial ID number is available. In both datasets, individual patient consent was signed for the anonymized use of the imaging and treatment planning data for retrospective analysis. These studies were conducted in accordance with the principles embodied in the Declaration of Helsinki.